\def\journal{prx}
\documentclass[aps,journal,amsmath,amssymb,reprint]{revtex4-1}
\usepackage{hyperref}
\usepackage{amsmath,amssymb}
\usepackage{graphicx}
\usepackage{overpic}
\usepackage{dashbox}
\usepackage{afterpage}
\usepackage[dvipsnames]{xcolor}
\usepackage[caption=false]{subfig}
\usepackage{MnSymbol}
\usepackage{mathtools}
\usepackage[english]{babel}
\usepackage{etoolbox}
\def\journal{prx}

\newcommand\phiinket{|\Phi_{\textrm{in}}\rangle}
\newcommand\phiinbra{\langle\Phi_{\textrm{in}}|}
\newcommand\phioutket{|\Phi_{\textrm{out}}\rangle}
\newcommand\phioutketetagamma{|\Phi_{\textrm{out}}^{\eta\gamma}\rangle}
\newcommand\phioutbraetagamma{\langle\Phi_{\textrm{out}}^{\eta\gamma}|}

\newcommand\DG{\langle\Delta\Gamma\rangle}

\newcommand\etagammain{|\eta \ \gamma\rangle^{\textrm{in}}}
\newcommand\etagammainbra{\prescript{\textrm{in}}{}{\langle} \gamma \ \eta|}
\newcommand\etagammaout{|\bareta \ \bargamma\rangle^{\textrm{out}}}

\newcommand{\bunderline}[1]{\underline{#1\mkern-4mu}\mkern4mu }

\newcommand\Smat{\bunderline{\bunderline{S}}}

\newcommand\Tr[1]{\textrm{Tr}\left\{#1\right\}}

\newcommand\bareta{\bar{\eta}}
\newcommand\bargamma{\bar{\gamma}}
\newcommand\Ttheta{T_{\Gamma}\left(\theta\right)}
\newcommand\Tmtheta{T_{\Gamma}\left(-\theta\right)}
\newcommand{\normF}[1]{\left\lVert#1\right\rVert_F}
\newcommand{\norm}[1]{\left\lVert#1\right\rVert}
\newcommand{\X}{X_{\bargamma\gamma}}
\newcommand{\zhat}{{\bf\hat{z}}}

\newcommand{\yhat}{{\bf\hat{y}}}

\begin{document}
\title{Observation-based symmetry breaking measures}
\author{Ivan Fernandez-Corbaton}
\email{ivan.fernandez-corbaton@kit.edu}
\affiliation{Institute of Nanotechnology, Karlsruhe Institute of Technology, 76021 Karlsruhe, Germany}
\begin{abstract}
Symmetry is one of the most general and useful concepts in physics. A system that has a symmetry is fundamentally constrained by it. The same constraints do not apply when the symmetry is broken. The quantitative determination of {\em how much a system breaks a symmetry} allows to reach beyond this binary situation and is a necessary step towards the quantitative connection between symmetry breaking and its effects. We introduce measures of symmetry breaking for a system interacting with external fields (particles). They can be computed from measurements of the system-mediated coupling strengths between subspaces of incoming and outgoing fields (particles) that transform in a definite way under the symmetry. The generality of these symmetry breaking measures and their tight connection to experimental measurements make them applicable to a very wide range of physics, like quantification of phase transitions, constraints in dynamical evolution, and the search for hidden symmetries.
\end{abstract}
\maketitle
\section{Introduction, summary and outline}
The study of symmetry is central in physics. At the most fundamental level, broken and unbroken symmetries guide the theories that explain and predict observations at all length scales: From elementary particle, nuclear and atomic scales \cite{Wigner1939,Nambu1961,Higgs1964}, through condensed matter \cite{Anderson1961,Wilson1983}, to cosmology \cite{Penrose2017}. When instead of the underlying theories, one considers a physical system like a molecule or a ferromagnetic crystal, the symmetries that the system has or lacks are crucial for understanding and predicting its properties and behavior. Examples of this are spectroscopy \cite{Wigner1959}, phase transitions \cite[Chap. 2]{Anderson1997}, and molecular chirality \cite{Barron2004}. Indeed, symmetry considerations are among the most general statements that we can make about physical systems. When an effect observed in a particular system is explained using only symmetry arguments, like ``the effect happens when the system has symmetry X and lacks symmetry Y'', the explanation is then valid and predictive in general: Systems that are vastly different from the original one will also exhibit the effect as long as they meet the symmetry requirements.

Given a system and a symmetry, the question of whether the system is symmetric leads to a binary outcome. A positive answer implies quantitative constrains like conservation laws \cite{Noether1918}. A negative answer precludes the same constraints. Broken symmetries are often only qualitatively considered, like in the explanation of ferromagnetic phase transitions or regarding the chirality of biomolecules. The question: {\em How much does the system break the symmetry?} is the starting point for quantitatively connecting symmetry breaking and its consequences. The question is addressed by symmetry breaking measures. Some measures have been defined for specific purposes like understanding spectra of nuclei and atoms \cite{French1967,Brody1981} and searching for hidden symmetries in the apparent disorder of liquids and colloidal glasses \cite{Wochner2009,Reichert2000}.  Very recently, the study of different symmetry breaking measures in a more general sense has shown their usefulness in quantifying quantum resources, studying quantum state evolution and estimation, analyzing accidental degeneracies, and quantifying spontaneous symmetry breaking \cite{Marvian2012,Marvian2014,Fang2016,Marvian2016,Dong2016,Dong2017}. These works already show large potential benefits in very diverse areas. Due to the generality of symmetry considerations we can reasonably expect many more areas of application, including currently unforeseen ones.

In this article we introduce measures of symmetry breaking for a system interacting with external fields (particles). Given a system and a symmetry, the corresponding measure can be computed from the system-mediated coupling strengths between subspaces of incoming and outgoing fields (particles) that transform in a definite way under the symmetry. These coupling strengths can be experimentally determined by amplitude squared measurements, that is, by measuring particle numbers of quantum states or classical field strength squares. This links the symmetry breaking measures to general measurement theory \cite{Sorkin1994,Hardy2001}. We will refer to amplitude squared measurements as {\em intensity} measurements from now on. Neither a priori knowledge of the scattering operator of the system nor its experimental determination is needed to obtain the symmetry breaking measures. The details of the interaction do not need to be known either, which goes along well with the fact that symmetry considerations are valid independently of those details. We provide both formal and operational definitions of the symmetry breaking measures. In the case of continuous symmetries, the measures depend on a real valued parameter, like the rotation angle in rotations or the displacement in translations. We show how the parameter-independent intensity measurements allow to evaluate the breaking of the symmetry in a {\em global} way, i.e. for any value of the parameter. This becomes useful for uncovering hidden symmetries. Additionally, we show that when the interaction is unitary, the lowest term in the expansion of the measure with respect to a small value of the parameter, i.e. the {\em local} symmetry breaking property, measures the general ability of the system to exchange a conserved quantity with the incoming fields (particles), and provides an upper bound for the efficiency of such exchange, thereby constraining the dynamical evolution of the system. We start in Sec. \ref{sec:setting} by describing the formal setting that we will use and comment on its applicability. Section \ref{sec:quant} contains the main part of the results for continuous and discrete symmetries, together with the link to measurement theory. Examples are provided in Sec. \ref{sec:examples}. The {\em global} and {\em local} symmetry breaking properties are discussed in Sec. \ref{sec:discrete} before the discussion and conclusion section. 

\section{Setting and scope of applicability\label{sec:setting}}
\begin{figure}[h!]
	\includegraphics[width=\linewidth]{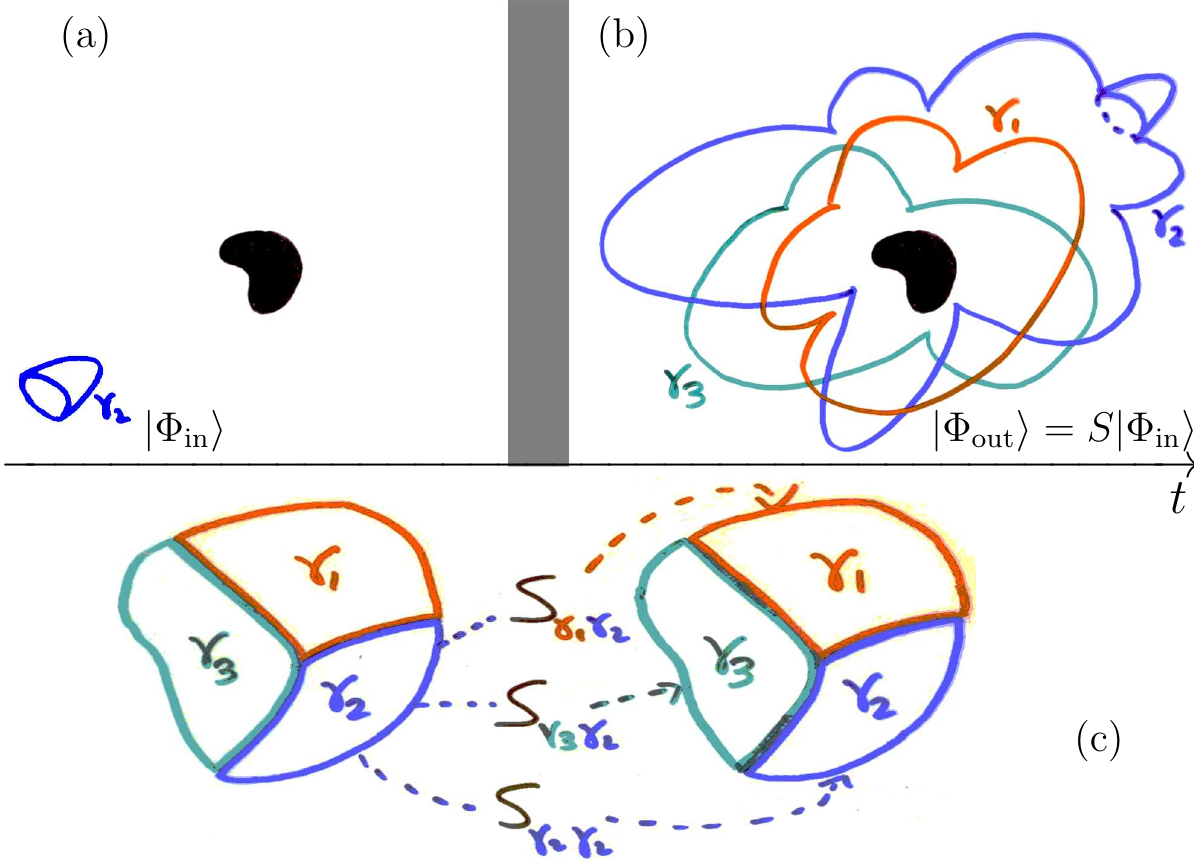}
	\caption{ \label{fig:byhand}(a) An incoming state $\phiinket$ interacts with a target system during a finite amount of time (gray region). (b) The interaction produces an outgoing state $\phioutket=S\phiinket$, where $S$ is the scattering operator of the system. (c) Given a symmetry, both incoming and outgoing Hilbert spaces can be partitioned into orthogonal subspaces characterized by the eigenvalues $\gamma$ of an operator determined by the symmetry. The restrictions $S_{\bargamma\gamma}$ of the scattering operator $S$ connect the incoming $\gamma$-subspace with the outgoing $\bargamma$-subspace. The coupling strengths $\X=\Tr{S_{\bargamma\gamma}^{\dagger}S_{\bargamma\gamma}}$ are measurable quantities which determine how strongly does the system break the symmetry [see Eqs. (\ref{eq:Mexp}) and (\ref{eq:Md}) for continuous and discrete symmetries, respectively].} 
\end{figure}

In the scattering operator setting \cite{Reed1979,Colton2012,Newton2013} illustrated in Fig. \ref{fig:byhand}, an incoming state interacts with the target system for a finite amount of time and produces an outgoing state. The states are represented by vectors in the incoming and outgoing Hilbert spaces of solutions of their dynamic equations in the absence of the target. The vectors of complete orthogonal basis for incoming and outgoing states are characterized by the eigenvalues of commuting operators. In such setting, the relevant information about the target is its scattering operator $S$: A linear and bounded mapping that the target mediates between the incoming and outgoing spaces. A symmetry transformation is represented by a unitary operator acting on the vectors in these spaces, which can also be used to transform other operators. There are two kinds of symmetries, discrete like parity, and continuous like translations. The continuous transformations are generated by a Hermitian operator, e.g: Linear momentum generates translations, and angular momentum generates rotations. 
The setting is general enough to cover a very wide range of cases: Classical or quantum, single or multi-particle scattering, with or without losses. The details will be different in each case. The Hilbert spaces could be simple, like in single particle or classical field scattering, or composed by direct sums of product spaces, like in multi-particle scattering where the number of particles is not necessarily conserved \cite[Chap. 2]{Altland2010}. The representations of operators will also change as exemplified by the different rotation matrices for fields (particles) of different spin.

It should be noted that this setting does not directly apply to the situation where there is no target system, but rather different incoming beams whose interaction results in the outgoing beams. In this case, the relevant scattering operator is the one due to the underlying theory of interacting fields (particles), like in quantum field theory \cite{Coleman1967}\cite[Chap. 3]{Weinberg1995}. Therefore, our focus here is on the quantification of the breaking of symmetries by target systems, and not of the breaking of symmetries by the underlying theories. 

\begin{figure}[h!]
	\includegraphics[width=0.85\linewidth]{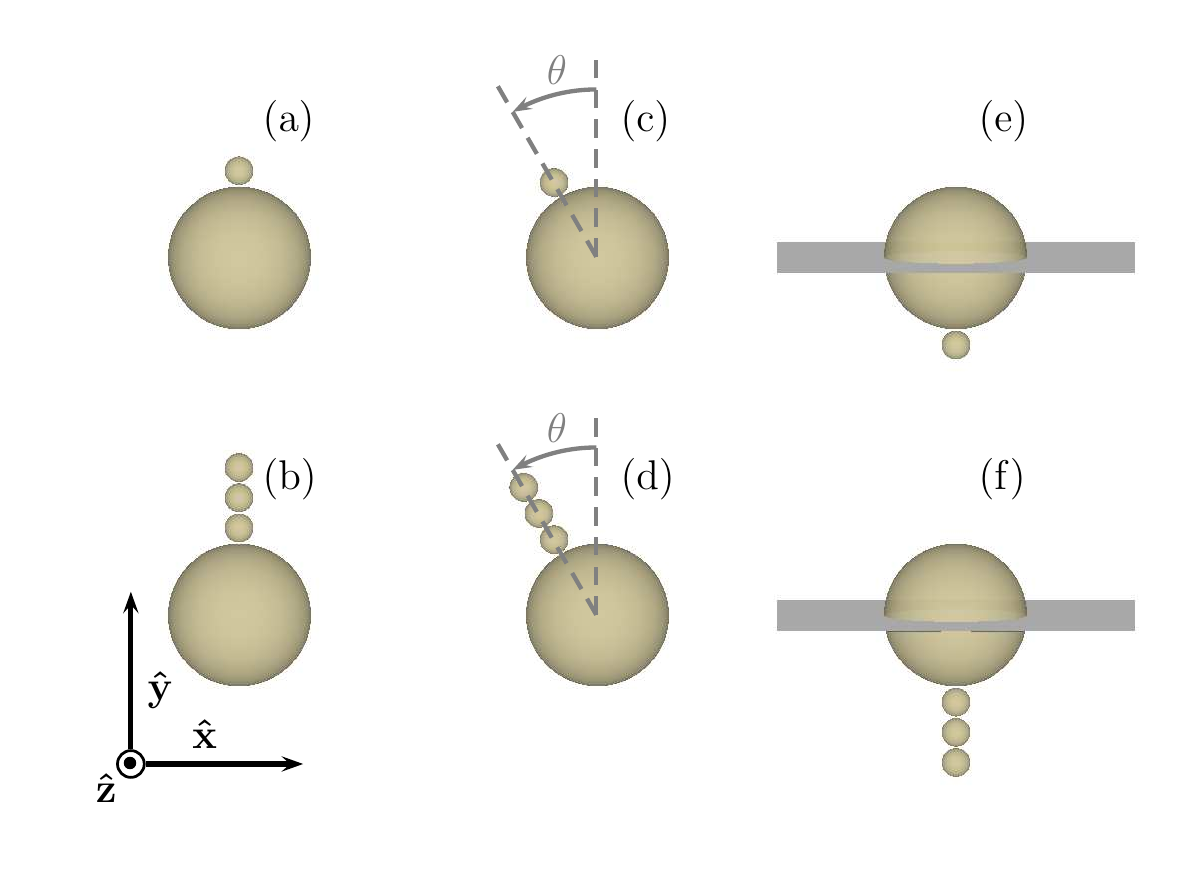}
	\caption{ \label{fig:spheres}Many of the rotational and mirror reflection symmetries of a sphere are broken by placing either one (a) or three (b) smaller spheres on top of it. While the symmetries are broken in both cases, intuition suggests that system (b) breaks them ``more'' than system (a). Panels (c) and (e) show the rotation and mirror reflection of system (a). Panels (d) and (f) show the rotation and mirror reflection of system (b). }
\end{figure}
\section{Quantification of symmetry breaking\label{sec:quant}}
Let us now exercise our intuition on the idea of quantifying broken symmetries. On the one hand, in Fig. \ref{fig:spheres}(a) we consider a sphere whose rotational symmetry is broken by a smaller sphere placed on top of it. On the other hand, in Fig. \ref{fig:spheres}(b), the symmetry is broken by stacking three small spheres on top of the larger sphere. Both systems are non-symmetric under rotations along the $\zhat$ axis, but we intuitively tend to consider that the second system ``breaks the symmetry more'' than the first because its asymmetric part is larger [see Figs. \ref{fig:spheres}(c) and \ref{fig:spheres}(d)]. Similarly for the mirror reflection across the $XZ$ plane seen in Figs. \ref{fig:spheres}(e) and \ref{fig:spheres}(f). In order to transfer these intuitions onto the previously described formal setting we start with the way in which an operator $O$ is transformed by a symmetry transformation represented by the unitary operator $T$: $O\rightarrow TOT^{-1}$. The basic idea for measuring how much does a scattering operator $S$ break a symmetry is to compare $S$ with its transformed version $TST^{-1}$. If we wanted to compare two three-vectors $\mathbf{a}$ and $\mathbf{b}$, we could use the squared Euclidean norm of its difference to compute the real non-negative number
\begin{equation}
	\label{eq:normvec}
	\frac{1}{2}\frac{|\mathbf{a}-\mathbf{b}|^2}{|\mathbf{a}|^2+|\mathbf{b}|^2},
\end{equation}
which is $0$ iff $\mathbf{a}=\mathbf{b}$, and is upper bounded by 1. The bound is reached when $\mathbf{a}=-\mathbf{b}$. The same basic idea can be used to compare two operators $A$ and $B$:
\begin{equation}
	\label{eq:normop}
\frac{1}{2}\frac{\norm{A-B}^2}{\norm{A}^2+\norm{B}^2}.
\end{equation}
The choice of {\em which} operator norm $\norm{\cdot}$ to use is not obvious\footnote{Even the choice of the Euclidean norm in the three-vector case above is arbitrary.}. We choose the Frobenius (Hilbert-Schmidt) operator norm $\normF{C}=\sqrt{\Tr{C^\dagger C}}$, where $\Tr{D}$ is the trace of $D$. We hence define the measure $M(S,T)$ of the breaking of symmetry $T$ by the system represented by $S$ as:
\begin{eqnarray}
	M\left(S,T\right)&=&\frac{1}{2}\frac{\normF{S-TST^{-1}}^2}{\normF{S}^2+\normF{TST^{-1}}^2}\\
												 &=&\frac{1}{2}\frac{\Tr{\left(S-TST^{-1}\right)^\dagger\left(S-TST^{-1}\right)}}{\Tr{S^\dagger S}+\Tr{\left(TST^{-1}\right)^\dagger TST^{-1}}}\nonumber\\
									 &=&\boxed{\frac{\Tr{\left(S-TST^{-1}\right)^\dagger\left(S-TST^{-1}\right)}}{4\Tr{S^\dagger S}}},\nonumber
\end{eqnarray}
where the last equality follows because a transformation by a unitary operator does not change the Frobenius norm. $M\left(S,T\right)$ is dimensionless, and takes values in $[0,1]$. The value $0$ means perfect symmetry. We note that the upper bound of $M$ is reached when $T$ and $S$ anti-commute ($T S=-ST$).

There is an important reason for our choice of operator norm.  At first sight, knowledge of $S$ seems to be required\footnote{If we know $S$, the choice of operator norm is not critical from the operational point of view because we can compute any operator norm we desire.} for computing $M\left(S,T\right)$. The crucial point about choosing the Frobenius norm is that, as later shown, it allows to obtain the measure of symmetry breaking from a reduced set of measurements of the outgoing intensity in scattering situations\footnote{This is a notable simplification with respect to the task of obtaining $S$ by experimental means. In general, both amplitude and phase measurements for complete sets of incoming states and outgoing measurements are needed to determine the complex elements of $S$. Phase measurements are known to be particularly challenging in many situations \protect\cite{Teague1983,Phillips2000}. $S$ may be analytically and/or numerically obtained when a model of the response of the system is available. Then, the experimentally obtained symmetry breaking measures allow to test the model.}. We now discuss the cases of continuous and discrete symmetries.
\subsection{Continuous symmetries}
A continuous symmetry is generated by a Hermitian operator $\Gamma$: $\Ttheta =\exp\left(-i\theta\Gamma\right)$, with $\Gamma^{\dagger}=\Gamma$, and the transformation depends on a real parameter $\theta$. For example, in the case of rotations $\theta$ is the rotation angle, and in the case of translations $\theta$ is the displacement. Then, the symmetry breaking measure depends on $\theta$: 
\begin{eqnarray}
	\label{eq:M}
		&& M\left[S,\Ttheta \right]=\\ 
		&&\frac{\Tr{\left[S-\Ttheta S\Tmtheta\right]^\dagger\left[S-\Ttheta S\Tmtheta\right]}}{4\Tr{S^\dagger S}},\nonumber
\end{eqnarray}
where we have used that ${\Ttheta}^{-1}=\Tmtheta$ because $\Ttheta$ is unitary\footnote{Then: ${\Ttheta}^{-1}=\left({\Ttheta}\right)^{\dagger}=\left[\exp\left(-i\theta\Gamma\right)\right]^{\dagger}=\exp\left(i\theta\Gamma\right)=\Tmtheta$}. The eigenvalues and eigenvectors of $\Gamma$ play an important role in what follows. The eigenvalues of $\Gamma$, which we denote by $\gamma$, are real numbers, and eigenvectors with different eigenvalue are orthogonal. This allows to decompose the incoming and outgoing spaces into orthogonal subspaces characterized by the different values of $\gamma$ [see Fig. \ref{fig:byhand}(c)]. We denote them $\gamma$-subspaces. They are typically multidimensional. As stated above, knowledge of $S$ is not needed to compute the measure $M$ of symmetry breaking in Eq. (\ref{eq:M}). $M$ can be obtained by illuminating the target system with eigenvectors of $\Gamma$ and measuring the total intensity in each outgoing $\bargamma$-subspace. Appendix \ref{app:continuous} shows that Eq. (\ref{eq:M}) can be written as:
{\small
	\begin{eqnarray}
	\label{eq:Mexp}
	&&M\left[S,\Ttheta \right]=\\
 &&\boxed{\frac{1}{4\left(\sum\limits_{\gamma,\bargamma}\X\right)}\sum_{\substack{p,q,n,m\\(p,q)\neq (0,0)\\(n,m)\neq (0,0)}}\frac{\left(i^{p-q+m-n}\right)\theta^{p+q+n+m}}{p!\ q!\ n!\ m!}\sum\limits_{\gamma,\bargamma}\gamma^{q+m}(\bargamma)^{p+n}\X}\nonumber.
\end{eqnarray}
}
where both $p-q+m-n$ and $p+q+n+m$ must be even. The $\X$ in Eq (\ref{eq:Mexp}) are the real valued coupling strengths that the system mediates from the subspace of incoming states with eigenvalue $\gamma$ to the subspace of outgoing states with eigenvalue $\bargamma$ [see Fig. \ref{fig:byhand}(c) and its caption]. We can write the $\X$ as (see App. \ref{app:experiment}):
\begin{equation}
 \label{eq:XX}
 \X=\normF{S_{\bargamma\gamma}}^2=\Tr{S_{\bargamma\gamma}^{\dagger}{S_{\bargamma\gamma}}},
 \end{equation}
 where $S_{\bargamma\gamma}$ are restrictions of the scattering operator $S$. As illustrated in Fig. \ref{fig:byhand}(c), they connect the incoming $\gamma$-subspace with the outgoing $\bargamma$-subspace. In a matrix representation, their matrix elements are the blocks that compose the total scattering matrix
\begin{equation}
	\protect\Smat=\begin{bmatrix}
		\protect\Smat_{\gamma_1\gamma_1}&\protect\Smat_{\gamma_1\gamma_2}&\protect\Smat_{\gamma_1\gamma_3}&\ldots\\
		\protect\Smat_{\gamma_2\gamma_1}&\protect\Smat_{\gamma_2\gamma_2}&\protect\Smat_{\gamma_2\gamma_3}&\ldots\\
		\vdots&\vdots&\vdots&\ddots
		\end{bmatrix}.
\end{equation}
\subsection{Discrete symmetries}
In the case of discrete symmetries, the role played by the eigenvectors and eigenvalues of $\Gamma$ in the continuous case is played by the eigenvectors and eigenvalues of the symmetry transformation $T$ itself. One difference is that the eigenvalues of $T$ do not need to be real. In App. \ref{app:discrete} we show that for a discrete symmetry $T$:
\begin{equation}
	\label{eq:Md}
	\boxed{M\left[S,T \right]=\frac{\sum\limits_{\gamma,\bargamma}\left[1-\mathbb{R}\left\{\gamma(\bargamma)^*\right\}\right]\X}{2\sum\limits_{\gamma,\bargamma}\X}},
\end{equation}
where $\mathbb{R}\{\cdot\}$ denotes the real part. 

Equations (\ref{eq:Mexp}) and (\ref{eq:Md}) can be considered operational definitions of continuous\footnote{It is worth noting that the number of $\Gamma$ eigenvalues can be infinite. For example, when considering rotations, $\gamma$ and $\bargamma$ can take any integer value. In theory this implies that an infinite number of $\X$ has to be obtained. In practice, the response of a system of finite size will eventually fall rapidly as the modulo of the angular momentum eigenvalue increases. This provides a limit to the number of needed $\X$ for the rotation case. Similar arguments can be found in other cases.} and discrete symmetry breaking measures, respectively. We highlight that their validity is independent of the details of the interaction, which are confined to the gray area in Fig. \ref{fig:byhand}. In App. \ref{app:experiment} we discuss briefly some generalities of the experimental measurement of $\X$.
\subsection{The Frobenius norm link to measurement theory}
Further analysis of the key quantities $\X$ ties the symmetry breaking measures to generalized measurement theory \cite{Sorkin1994,Hardy2001}. The $\X$ can also be written as (see App. \ref{app:experiment})
\begin{equation}
 \label{eq:XXX}
 \X=\sum_{\eta}\sum_{\bareta}\left|\beta^{\eta\gamma}(\bareta,\bargamma)\right|^2,
 \end{equation}
 where $\beta^{\eta\gamma}(\bareta,\bargamma)$ are the coordinates of the outgoing state produced by the system upon interaction with the incoming state $\etagammain$. The $\etagammain$ are the members of an orthonormal basis of incoming states. The coordinates $\beta^{\eta\gamma}(\bareta,\bargamma)$ refer to a similar outgoing basis with members $|\bareta\ \bargamma\rangle^{\text{out}}$. The $\eta$ and $\bareta$ labels are composite indexes that complete the characterization of the basis states. Equation (\ref{eq:XXX}) shows that the $\X$, and hence the symmetry breaking measures, are directly determined by the modulus squared of complex amplitudes $\left(\left|\beta^{\eta\gamma}(\bareta,\bargamma)\right|^2\right)$. When the theory of measurement is minimally extended beyond classical probability theory in order to explain interference effects, it is precisely the squares of complex amplitudes which arise naturally and provide the origin of the trace-rule for computing the probabilities of measurement outcomes. The extensions encompass quantum mechanics \cite{Hardy2001} and generalizations thereof \cite{Sorkin1994}. We take this link as a vindication of our (and that of others \cite{Fang2016}) choice of the Frobenius norm.
 
\section{Examples\label{sec:examples}}
Before providing numerical examples, we place some aspects of two prominent phenomena due to symmetry breaking in the context provided so far. First, the vacuum expectation value of the Higgs boson is a measure of the electroweak symmetry breaking, and can be interpreted as the vacuum-mediated coupling of fermions with different eigenvalues of the chirality operator. And second, the ferromagnetic phase transition is related to the breaking of rotational symmetry caused by the alignment of electron spins. In this case, the relevant $\X$ to be measured are the couplings between subspaces of different angular momenta. Then, the corresponding symmetry breaking measure of Eq. (\ref{eq:Mexp}) can be seen as a degree of phase transition, as proposed in \cite{Dong2016} with a different symmetry breaking measure.

\begin{figure}[h!]
	\includegraphics[width=1.0\linewidth]{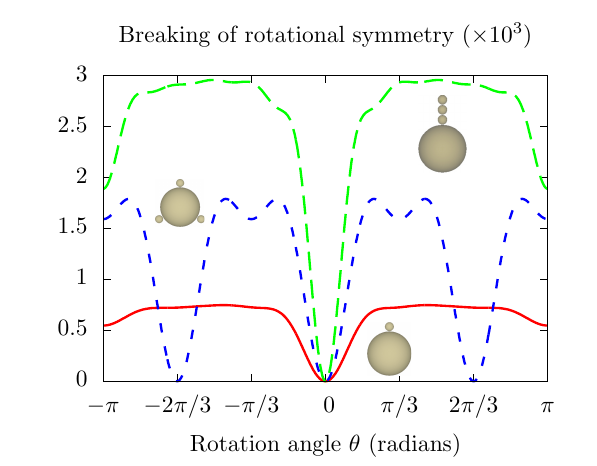}
	\caption{ \label{fig:breakings} Dimensionless measure of the breaking of rotational symmetry by each of the three systems displayed inside the figure as a function of the rotation angle $\theta$. The measure is computed using Eq. (\ref{eq:M}) particularized to rotations. For each system, the rotation axis is centered in the middle of the larger sphere and is perpendicular to the plane of the paper. The green long-dashed line corresponds to the system in Fig. \ref{fig:spheres}(b), and is always above the continuous red line corresponding to the system in Fig. \ref{fig:spheres}(a). This agrees with our initial intuition in this respect (see the beginning of Sec. \ref{sec:quant}), and allows to {\em quantify} it. The zeros at $\theta=\pm 2\pi/3$ of the blue short-dashed line reflect the discrete rotational symmetry of the corresponding system (see Sec. \ref{sec:discrete}).}
\end{figure}

Let us now go back to the two systems of Fig. \ref{fig:spheres} in the context of classical electromagnetic scattering and provide examples of theoretically computed symmetry breaking measures. For simplicity, we consider a monochromatic excitation. Appendix \ref{app:numerical} contains the description of the numerical calculations. We remark that while we choose the particular case of classical electromagnetic scattering for convenience, the ideas and results contained in the article apply to the general case in the scattering setting. Figure \ref{fig:breakings} shows the angle-dependent breaking of rotational symmetry along the $\zhat$ axis. The continuous red line corresponds to the system in Fig. \ref{fig:spheres}(a), and the long-dashed green line to the system in Fig. \ref{fig:spheres}(b). 
The short-dashed blue line will be discussed later. Our initial intuition is reflected and {\em quantified} by the results: The system with three small spheres breaks rotational symmetry more strongly than the system with one small sphere for all $\theta$. We now consider the mirror symmetry across the $XZ$ plane shown in Fig. \ref{fig:spheres}(e) and \ref{fig:spheres}(f). The breaking of the mirror symmetry across the $XZ$ plane is $5.46\times 10^{-4}$ for the system in Fig. \ref{fig:spheres}(a) and $1.89\times 10^{-3}$ for the system in Fig. \ref{fig:spheres}(b), again confirming and {\em quantifying} our intuition. The results also show a reassuring mutual consistency between the measures of the breaking of different symmetries in the same system. For each of the two systems, the above given numbers for the breaking of the mirror symmetry coincide exactly with the values of rotational symmetry breaking for $\theta=\pi$ in Fig. \ref{fig:breakings}. Indeed, in these two systems, the mirror symmetry and the rotation by $\pi$ along the $\zhat$ axis have the same effect.
\section{Global and local symmetry breaking\label{sec:discrete}}
The short-dashed blue line in Fig. \ref{fig:breakings} represents the rotational symmetry breaking of the system displayed next to it, where three small spheres are placed around a larger sphere in a way that achieves a three-fold discrete rotational symmetry. This symmetry is reflected in the zeros of the symmetry breaking function at $\theta=\pm 2\pi/3$. It is important to note that, for continuous symmetries, the knowledge of the $\X$ coupling strengths allows us to use Eq. (\ref{eq:Mexp}) for computing the symmetry breaking for any $\theta$. This {\em global} reach is very useful for uncovering ``hidden'' symmetries of the system, and could be exploited to improve current techniques \cite{Wochner2009,Reichert2000}. 

Let us now turn to the {\em local} properties of continuous symmetry breaking by a system. Roughly speaking, we are after the symmetry breaking at the onset of the transformation when $\theta\rightarrow 0$. It is straightforward to show (see App. \ref{app:last}) that, to lowest order in $\theta$, which is actually $\theta^2$
{\small
\begin{equation}
	\label{eq:Msmall}
				M\left[S,\Ttheta  \right]\approx
				\theta^2\boxed{\frac{\Tr{[S,\Gamma]^\dagger[S,\Gamma]}}{4\Tr{S^\dagger S}}}=\theta^2B_{\Gamma}.
\end{equation}
}
We can take the boxed expression as the $\theta$ independent ``slope'' of symmetry breaking. It is proportional to the Frobenius norm squared of the commutator between $S$ and the generator $\Gamma$. The value of $B_\Gamma$ can be obtained from the $\X$ coupling strengths with the formula 
\begin{equation}
	\label{eq:Msmalllab}
	B_{\Gamma}=\frac{\Tr{[S,\Gamma]^\dagger[S,\Gamma]}}{4\Tr{S^\dagger S}}=\frac{\sum\limits_{\gamma,\bargamma} \left(\gamma-\bargamma\right)^2\X}{4\sum_{\bargamma}^{\gamma}\X},
\end{equation}
which results from keeping only the terms containing $\theta^2$ in Eq. (\ref{eq:Mexp}). It can be shown (App. \ref{app:b0m0}) that when $B_{\Gamma}=0$ all other higher order terms also vanish and $M\left[S,\Ttheta  \right]=0$ for all $\theta$.

We will finish discussing a link with dynamical evolution. In the absence of absorption and gain, we show in App. \ref{app:last} that the expression 
\begin{equation}
	\label{eq:limit}
	C_{S\Gamma}=\sqrt{B_{\Gamma}\left(4\Tr{S^\dagger S}\right)}=\sqrt{\sum\limits_{\gamma,\bargamma} \left(\gamma-\bargamma\right)^2\X},
\end{equation}
provides i) a direct measure of the excitation independent ability of the system to exchange the quantity represented by $\Gamma$ with the incoming states, and ii) an upper bound for such exchange with any normalized incoming state. For example, exchanges of linear and angular momenta are responsibles for the forces and torques exerted by the external fields (particles) onto the system. These exchanges satisfy conservation laws ultimately due to symmetry. This, and previous \cite{Marvian2014} links, between symmetry breaking and dynamical constraints, together with recent extensions of Noether's theorem \cite{Marvian2014,FerCor2016c}, provide starting points for the precise quantitative connection between symmetry breaking and the joint evolution of the fields (particles) and target system.

\section{Discussion and conclusions}
Before finishing, we comment on the possibility of a particular extension of the ideas and results contained in this article. Modern studies of symmetry breaking in the context of quantum field theory are probing the invariance of our fundamental theories under discrete and continuous symmetries, like CPT and Lorentz invariance \cite{Kosteleck2004,Mattingly2005,Liberati2013}. These symmetries are assumed to hold in current models, but their breaking is predicted or at least allowed in some of the theories that ultimately aim to achieve the unification of gravity with the other known interactions. As previously stated, the scattering setting that we have used is adapted for quantifying the symmetry breaking of a target system, not of the underlying theory of interactions \cite{Coleman1967}\cite[Chap. 3]{Weinberg1995}. Nevertheless, the following extension seems plausible. In a target-less scenario we consider several incoming beams of fields (particles) that will interact and result in outgoing products. The mapping between incoming and outgoing is now provided by the $S_{th}$ operator due to the underlying theory of interactions. Conceptually, it seems plausible to use Eqs. (\ref{eq:Mexp}) or (\ref{eq:Md}) to quantify the breaking of a symmetry $T$ by the operator $S_{th}$ in the following way. When considered as a single entity, all the incoming beams together should be prepared as an eigenstate of $T$, and the cross-couplings $\X$ to outgoing subspaces with different eigenvalue should be measured [see Fig. \ref{fig:breakings}(c)]. Then, Eqs. (\ref{eq:Mexp}) or (\ref{eq:Md}) can be applied. 

In conclusion, we have introduced observation-based symmetry breaking measures that combine two very general concepts: Symmetry and intensity measurements. Besides applications in the spectroscopic determination of the symmetries of target systems, the generality of their definition and their tight connection to measurement make these symmetry breaking measures good candidates for establishing quantitative relations between symmetry breaking and its effects across a very wide range of physics, like for example in phase transitions and dynamic evolution. I believe that the full potential of the systematic quantification of symmetry breaking has not yet been established, and that it will be a very productive research area. 
\begin{acknowledgments}
The author acknowledges Ms. Magda Felo for her help with the figures, Prof. Carsten Rockstuhl for useful feedback on the manuscript, and the support by the Deutsche Forschungsgemeinschaft and the open access publishing fund of the Karlsruhe Institute of Technology.
\end{acknowledgments}

\ifdefstring{\journal}{prl}{\end{document}}{}
\ifdefstring{\journal}{prx}{\appendix\makeatletter{}\section{Measuring symmetry breaking}
\subsection{Continuous symmetry\label{app:continuous}}
We now prove that, for a continuous symmetry $\Ttheta=\exp\left(-i\theta\right)$, Eq. (\ref{eq:M}) can be written as Eq. (\ref{eq:Mexp}). First, we address its numerator. The expansion of the exponential

\begin{equation}
	\Ttheta=\exp\left(-i\theta\right)=\sum_{l=0}^\infty\frac{\left(-i\theta \Gamma\right)^l}{l!},
\end{equation}

allows to readily show the following equality:

\begin{equation}
	\label{eq:one}
	\begin{split}
		&S-\Ttheta S\Tmtheta =S-\sum^{\infty}_{n=0}\sum^{\infty}_{m=0} \frac{\left(-i\theta\right)^n}{n!}\frac{\left(i\theta\right)^m}{m!}\Gamma^nS\Gamma^m\\
		&=\sum_{\substack{n,m\\(n,m)\neq (0,0)}}\frac{\left(i^{m-n}\right)\theta^{n+m}}{n!\ m!}\Gamma^nS\Gamma^m,
	\end{split}
\end{equation}

which we then use twice to show that the operator inside the trace in the numerator of Eq. (\ref{eq:M}) can be written as:

\begin{equation}
	\label{eq:two}
	\begin{split}
		&\left[S-\Ttheta S\Tmtheta \right]^\dagger\left[S-\Ttheta S\Tmtheta \right]\\
		&=\sum_{\substack{p,q\\(p,q)\neq (0,0)}}\sum_{\substack{n,m\\(n,m)\neq (0,0)}}\frac{\left(i^{p-q+m-n}\right)\theta^{p+q+n+m}}{p!\ q!\ n!\ m!}\Gamma^qS^\dagger\Gamma^{p+n}S\Gamma^m,
	\end{split}
\end{equation}
where we also use the hermiticity of any power of $\Gamma$: $\left(\Gamma^s\right)^\dagger=\Gamma^s$.

The trace of any operator $O$ is a basis independent quantity. In particular, we can use a basis of incoming states where all its vectors are eigenvectors of $\Gamma$: $\Gamma\etagammain=\gamma\etagammain$ for real $\gamma$, to write

\begin{equation}
\label{eq:tr}
	\Tr{O}=\sum_{\eta,\gamma}\etagammainbra O\etagammain,
\end{equation}
where $\eta$ is a composite index collecting the rest of the eigenvalues characterizing each basis vector, and the summation index represents sums over discrete eigenvalues and integrals over continuous ones.

The trace of the operator of Eq. (\ref{eq:two}) is the weighted sum of the traces of the operators $\Gamma^qS^\dagger\Gamma^{p+n}S\Gamma^m$, which, using Eq. (\ref{eq:tr}) we write as
\begin{equation}
	\label{eq:numbers}
	\begin{split}
		\Tr{\Gamma^qS^\dagger\Gamma^{p+n}S\Gamma^m}&=\sum_{\eta,\gamma}\etagammainbra\Gamma^qS^\dagger\Gamma^{p+n}S\Gamma^m\etagammain\\
																				 &=\sum_{\eta,\gamma}\left(\gamma^{q+m}\right)\etagammainbra S^\dagger\Gamma^{p+n}S\etagammain\\
		&=\sum_{\eta,\gamma}\gamma^{q+m}\phioutbraetagamma \Gamma^{p+n} \phioutketetagamma,
	\end{split}
\end{equation}
where for the second equality we use that the $\etagammain$ are eigenvectors of $\Gamma$ with real eigenvalues $\gamma$, and the third contains the definition of $\phioutketetagamma=S\etagammain$ as the outgoing state corresponding to an incoming $\etagammain$. 

It is clear from Eq. (\ref{eq:numbers}) that the numerator of Eq. (\ref{eq:M}) can be obtained from the collection of real numbers $\gamma^{q+m}\phioutbraetagamma \Gamma^{p+n} \phioutketetagamma$. Let us now consider one of them. 

We use the coordinates of the outgoing state in an outgoing basis of eigenstates of $\Gamma$ $\{\etagammaout\}$ 

\begin{equation}
	\label{eq:defi}
	\phioutketetagamma=S\etagammain=\sum_{\bareta,\bargamma}\beta^{\eta\gamma}(\bareta,\bargamma)\etagammaout,
\end{equation}

and write

\begin{equation}
	\label{eq:oneterm}
	\begin{split}
		&\gamma^{q+m}\phioutbraetagamma \Gamma^{p+n} \phioutketetagamma=\gamma^{q+m}\sum_{\bareta,\bargamma}(\bargamma)^{p+n}|\beta^{\eta\gamma}(\bareta,\bargamma)|^2\\
		&=\gamma^{q+m}\sum_{\bargamma}(\bargamma)^{p+n}X^{\eta\gamma}_{\bargamma},
	\end{split}
\end{equation}
where in the first equality we use that $\Gamma^{s}$ is diagonal in the $\{\etagammaout\}$ basis, with elements $\bargamma^s$, and the second one implicitly defines $X^{\eta\gamma}_{\bargamma}=\sum_{\bareta}|\beta^{\eta\gamma}(\bareta,\bargamma)|^2$. That is, $X^{\eta\gamma}_{\bargamma}$ is the portion of the total Euclidean squared norm of the outgoing state $\phioutketetagamma$ contained in the $\bargamma$-subspace. We now plug the result of Eq. (\ref{eq:oneterm}) back into Eq. (\ref{eq:numbers}), and manipulate it to obtain:
	\begin{equation}
		\begin{split}
			\Tr{\Gamma^qS^\dagger\Gamma^{p+n}S\Gamma^m}&=\sum_{\eta,\gamma}\gamma^{q+m}\phioutbraetagamma \Gamma^{p+n} \phioutketetagamma\\
												   &=\sum_{\eta,\gamma}\gamma^{q+m}\sum_{\bargamma}(\bargamma)^{p+n}X^{\eta\gamma}_{\bargamma}\\
						&=\sum_{\gamma,\bargamma}\gamma^{q+m}(\bargamma)^{p+n}\sum_{\eta}X^{\eta\gamma}_{\bargamma}\\
						&=\sum_{\gamma,\bargamma}\gamma^{q+m}(\bargamma)^{p+n}\X,
		\end{split}
\end{equation}
where the last equality defines $\sum_{\eta}X^{\eta,\gamma}_{\bargamma}=\X$. That is, the outgoing intensity accumulated in the $\bargamma$-subspace upon separate excitation of the system by the $\etagammain$ incoming states for fixed $\gamma$ and all $\eta$.

We now treat the denominator of  Eq. (\ref{eq:M}):
\begin{equation}
	\label{eq:deno}
	\begin{split}
		4\Tr{S^\dagger S}&=4\sum_{\eta,\gamma} \etagammainbra S^\dagger S\etagammain=4\sum_{\eta,\gamma}\phioutbraetagamma\Phi_{\text{out}}^{\eta\gamma}\rangle\\
		&=4\sum_{\eta,\gamma}\sum_{\bargamma}X_{\bargamma}^{\eta\gamma}=4\sum_{\gamma,\bargamma}\X.
	\end{split}
\end{equation}
Equation (\ref{eq:M}) can finally be written as:
{\small
	\begin{eqnarray}
	\label{eqapp:Mexp}
	&&M\left[S,\Ttheta \right]=\\
 &&\boxed{\frac{1}{4\left(\sum\limits_{\gamma,\bargamma}\X\right)}\sum_{\substack{p,q,n,m\\(p,q)\neq (0,0)\\(n,m)\neq (0,0)}}\frac{\left(i^{p-q+m-n}\right)\theta^{p+q+n+m}}{p!\ q!\ n!\ m!}\sum\limits_{\gamma,\bargamma}\gamma^{q+m}(\bargamma)^{p+n}\X}\nonumber.
\end{eqnarray}
}
We note that there are restrictions in the possible values of $(p,q,n,m)$ besides the ones indicated in the formula. Since $M[S,\Ttheta]$ is real, we can restrict the exponent in $i^{p-q+m-n}$ to be even. Also, $M[S,\Ttheta]$ is an even function of $\theta$. This is readily shown from Eq. (\ref{eq:M}) using that $\Tmtheta={\Ttheta}^{-1}$ and the cyclic property of the trace $\Tr{ABC}=\Tr{BCA}$. This restricts the exponent in $\theta^{p+q+n+m}$ to also be even. Equation (\ref{eqapp:Mexp}) can be considered an operational definition of the $\theta$-dependent breaking of a continuous symmetry.
\subsection{Discrete symmetry\label{app:discrete}}
In order to prove Eq. (\ref{eq:Md}), we start by writing its numerator as
\begin{equation}
	\label{eq:somi}
	\begin{split}
		&\Tr{\left[S-TST^{-1}\right]^\dagger\left[S-TST^{-1}\right]}\\
		&=\Tr{S^\dagger S-S^\dagger T S T^{-1}-{T^{-1}}^\dagger S^\dagger T^\dagger S+{T^{-1}}^\dagger S^\dagger T^\dagger TST^{-1}}\\
		&=2\Tr{S^\dagger S}-\Tr{S^\dagger T S T^{\dagger}+TS^\dagger T^\dagger S},
	\end{split}
\end{equation}

where we have used the unitarity of  $T$: $T^\dagger=T^{-1},\ T^\dagger T=T T^\dagger=I$, and the cyclic property of the trace $\Tr{ABC}=\Tr{BCA}$.

The first term in the last line of Eq. (\ref{eq:somi}) is $2\Tr{S^\dagger S}=2\sum_{\gamma,\bargamma}\X$, as per Eq. (\ref{eq:deno}). We now work on the second term:

\begin{equation}
	\begin{split}
		&\Tr{TS^\dagger T^\dagger S+S^\dagger T S T^{\dagger}}\\
		&=\sum_{\eta,\gamma}\etagammainbra TS^\dagger T^\dagger S+S^\dagger T S T^{\dagger} \etagammain\\
		&=\sum_{\eta,\gamma}\left[\gamma \etagammainbra S^\dagger T^\dagger S \etagammain+\gamma^*\etagammainbra S^\dagger T S \etagammain\right]\\
		&	=\sum_{\eta,\gamma}\left[\gamma  \sum_{\bareta,\bargamma} {\bargamma}^*|\beta^{\eta\gamma}(\bareta,\bargamma)|^2+\gamma^*\sum_{\bareta,\bargamma} \bargamma|\beta^{\eta\gamma}(\bareta,\bargamma)|^2\right]\\
		& = \sum_{\gamma,\bargamma}\left({\bargamma}^*\gamma+\gamma\bargamma^*\right)\X,
	\end{split}
\end{equation}
where in the second equality we use $T^{\dagger}\etagammain=\gamma^*\etagammain$, $(\cdot)^*$ denoting complex conjugation, in the third we use Eq. (\ref{eq:defi}), and in the fourth the definition of $\X$ in App. \ref{app:continuous} or Eq. (\ref{eq:doit}).

Going back to Eq. (\ref{eq:somi}), the numerator of Eq. (\ref{eq:Md}) is then 
\begin{equation}
2\sum_{\gamma,\bargamma}\X-\sum_{\gamma,\bargamma}\left({\bargamma}^*\gamma+\gamma\bargamma^*\right)\X=2\sum_{\gamma,\bargamma}\left[1-\mathbb{R}\left\{\gamma(\bargamma)^*\right\}\right]\X,
\end{equation}
with which Eq. (\ref{eq:Md}) follows immediately.

\subsection{Measurement considerations\label{app:experiment}}
We now discuss some aspects of the experimental measurement of $\X$. We recall their definition from App. \ref{app:continuous}:

\begin{equation}
	\label{eq:doit}
	\X=\sum_{\eta}X_{\bargamma}^{\eta\gamma}=\sum_{\eta}\sum_{\bareta}\left|\beta^{\eta\gamma}(\bareta,\bargamma)\right|^2,
\end{equation}
where $\beta^{\eta\gamma}(\bareta,\bargamma)$ are the coordinates of the outgoing state upon excitation with the incoming $\etagammain$.

Operationally, we can fix $\gamma$ and $\bargamma$, successively excite the target with $\etagammain$ for all $\eta$, measure $\sum_{\bareta}\left|\beta^{\eta\gamma}(\bareta,\bargamma)\right|^2$ for each $\eta$, i.e. the outgoing intensity in the $\bargamma$-subspace, and finally add up the results for all $\eta$ to obtain $\X$. 

If $S_{\bargamma\gamma}$ denotes the restriction of the scattering operator to incoming states in the $\gamma$-subspace and outgoing states in the $\bargamma$-subspace, the above described procedure will yield:
\begin{equation}
	\label{eq:xappus}
	\X=\sum_{\eta}\etagammainbra S_{\bargamma\gamma}^{\dagger}{S_{\bargamma\gamma}}\etagammain=\Tr{S_{\bargamma\gamma}^{\dagger}{S_{\bargamma\gamma}}}=\normF{S_{\bargamma\gamma}}^2.
\end{equation}

The way to measure the intensity in an outgoing $\bargamma$-subspace depends on the particular situation. For example, if we want to measure the intensity in each of the two helicity subspaces in nano-optics, we may first collimate the field outgoing from a localized scatterer with helicity preserving lenses, and then use quarter wave plates, linear polarizers and Charged Coupled Device (CCD) cameras (see the setup in \cite[Fig.1]{Tischler2014} for an example that collects the forward transmitted outgoing field from a nanohole. A similar setup would also be needed in reflection). Simultaneous measurements of the intensity in several $\bargamma$-subspaces are also possible. The setup in \cite[Fig.1]{Tischler2014} can be extended with a polarizer beam-splitter after the quarter wave plate, and an extra CCD camera to measure the outgoing intensity in both $\pm$ helicity subspaces simultaneously. Similarly, simultaneous measurements of the intensity in different angular momentum subspaces can be performed \cite{Lavery2013,Ren2016}. 

Let us now turn to the excitation. In the previous discussion we have assumed that the illuminating field is a single $\etagammain$, and that the input values of $\eta$ and $\gamma$ are successively changed to obtain all the $\X$ by accumulation of the $X^{\eta,\gamma}_{\bargamma}$. This is not the only possible excitation strategy. For example, let us assume that the incoming state is a general $\Gamma$ eigenstate 
\begin{equation}
	\label{eq:x}
|\Phi_\gamma\rangle=\sum_{\eta}\alpha(\eta)\etagammain,
\end{equation}
where the $\alpha(\eta)$ are uncorrelated complex random variables with zero mean and equal variance $\sigma^2$. We will now show that, while the intensity in the outgoing $\bargamma$-subspace is then random, its average value is proportional to $\X$. The deviation from the average will be reduced as the output intensity measurements are integrated over more instances of the random incoming state. In optics, a spatial light modulator can be used to obtain the subsequent realizations of the random excitation.

The intensity in the outgoing $\bargamma$-subspace due to the excitation of the target with the random eigenstate of $\Gamma$ in Eq. (\ref{eq:x}) is
\begin{equation}
	P_{\bargamma\gamma}=\langle\Phi_\gamma|S_{\bargamma\gamma}^{\dagger}S_{\bargamma\gamma}|\Phi_\gamma\rangle,
\end{equation}
which is a random variable. Its average $\mathbb{E}\left\{P_{\bargamma\gamma}\right\}$ is $\sigma^2\X$: 

\begin{equation}
	\begin{split}
	\mathbb{E}\left\{\langle\Phi_{\gamma}|S_{\bargamma\gamma}^{\dagger}S_{\bargamma\gamma}|\Phi_{\gamma}\rangle\right\}&=\mathbb{E}\left\{\sum_{\hat{\eta},\tilde{\eta}} \alpha(\hat{\eta})^*\alpha(\tilde{\eta})\langle \gamma \ \hat{\eta}|S_{\bargamma\gamma}^{\dagger}S_{\bargamma\gamma}|\tilde{\eta}\ \gamma\rangle\right\}\\
	&=\sum_{\hat{\eta},\tilde{\eta}}\mathbb{E}\left\{\alpha(\hat{\eta})^*\alpha(\tilde{\eta})\right\}\langle \gamma \ \hat{\eta}|S_{\bargamma\gamma}^{\dagger}S_{\bargamma\gamma}|\tilde{\eta}\ \gamma\rangle\\
	&=\sum_{\hat{\eta}}\mathbb{E}\left\{|\alpha(\hat{\eta})^2|\right\}\langle \gamma \ \hat{\eta}|S_{\bargamma\gamma}^{\dagger}S_{\bargamma\gamma}|\hat{\eta}\ \gamma\rangle\\
	&=\sigma^2\sum_{\hat{\eta}}\langle \gamma \ \hat{\eta}|S_{\bargamma\gamma}^{\dagger}S_{\bargamma\gamma}|\hat{\eta}\ \gamma\rangle=\sigma^2\X,
	\end{split}
\end{equation}
where the third equality follows because the $\alpha(\eta)$ are uncorrelated and have zero mean, and the last one follows from Eq. (\ref{eq:xappus}). When using $P_{\bargamma\gamma}$ to compute Eqs. (\ref{eq:Mexp}) and (\ref{eq:Md}), the factor $\sigma^2$ will be present in both numerator and denominator and cancel out.

\section{Numerical calculations\label{app:numerical}}
The symmetry breaking measures shown Fig. \ref{fig:breakings} and given in the text are computed numerically by first obtaining the scattering matrix of the systems and then applying Eq. (\ref{eq:M}). The $S$-matrix of each system is readily computed using Mie theory and composite $T$-matrix techniques \cite{Xu1995}, which use the basis of multipolar fields truncated to a maximum multipolar order $j_{max}$. Dipolar interactions correspond to $j=1$, quadrupolar interactions to $j=2$, and so on. We use here $j_{max}=10$, beyond which we neglect the rest of the (very small) terms in the infinite dimensional scattering matrix. Indeed, in each of the three systems, the sum of all squared $j=10$ terms is less than a part in $10^5$ of the total sum of squared terms for $j=1\ldots 10$. After obtaining the $T$-matrix, the $S$-matrix is given by: $S=I+2T$, where $I$ is the identity. We have restricted the calculations to monochromatic excitations for simplicity. In units of wavelength, the radii of the big and small spheres in Fig. \ref{fig:spheres} and Fig. \ref{fig:breakings} are 0.5 and 0.1, respectively, and the smallest separation between them 0.01. The spheres are surrounded by vacuum and have a relative permittivity equal to 10. Once the $S$-matrix is known, the only piece missing for computing Eq. (\ref{eq:M}) is the matrix representation of the symmetry operators. The rotation matrices in the multipolar basis can be obtained for example from \cite[Eqs. 7.3-(15-17)]{Tung1985}, and the mirror symmetry matrices from multiplying the matrix representing a rotation by $\pi$ along the $\yhat$ axis with the matrix representing the parity operator, whose elements can be obtained from \cite[Eq. 11.4-7]{Tung1985}.

\section{The slope of symmetry breaking in conservation laws\label{app:last}}
In order to obtain the lowest order term of a measure of continuous symmetry breaking, we substitute $\Ttheta  S \Tmtheta $ by its approximation using $\Ttheta=\exp\left(-i\theta \right)\approx I-i\theta\Gamma$, and $\Tmtheta \approx I+i\theta\Gamma$. The difference between the original and transformed operators is then:
\begin{equation}
		\begin{split}
				&S-\Ttheta  S \Tmtheta \approx\\
				& S-(I-i\theta\Gamma)S(I+i\theta\Gamma) = -i\theta[S,\Gamma]-\theta^2\Gamma S\Gamma,
		\end{split}
\end{equation}
where $[S,\Gamma]=S\Gamma-\Gamma S$ is the commutator between $S$ and $\Gamma$. We keep only the first term, linear in $\theta$, and readily obtain from Eq. (\ref{eq:M})
{\small
\begin{equation}
	\label{eq:Msmall}
				M\left[S,\Ttheta  \right]\approx
				\theta^2\boxed{\frac{\Tr{[S,\Gamma]^\dagger[S,\Gamma]}}{4\Tr{S^\dagger S}}}=\theta^2B_{\Gamma}.
\end{equation}
}
We can take the boxed expression as the $\theta$ independent ``slope'' of symmetry breaking. As per Eq. (\ref{eq:Msmalllab}):
\begin{equation}
	B_{\Gamma}=\frac{\Tr{[S,\Gamma]^\dagger[S,\Gamma]}}{4\Tr{S^\dagger S}}=\frac{\sum\limits_{\gamma,\bargamma} \left(\gamma-\bargamma\right)^2\X}{4\sum_{\bargamma}^{\gamma}\X}.
\end{equation}

Going back to the systems in Fig. \ref{fig:spheres}, the slope of symmetry breaking for rotations along the $\zhat$ axis are: $B_{J_z}=4.06\times 10^{-3}$ for the system in Fig. \ref{fig:spheres}(a), and $B_{J_z}=2.92\times 10^{-2}$ for the system in Fig. \ref{fig:spheres}(b).

We now show that in the absence of absorption and gain, the expression 
\begin{equation}
	\label{eq:limit2}
	C_{S\Gamma}=\sqrt{B_{\Gamma}\left(4\Tr{S^\dagger S}\right)}=\sqrt{\sum\limits_{\gamma,\bargamma} \left(\gamma-\bargamma\right)^2\X},
\end{equation}
provides on the one hand a direct measure of the ability of the system to exchange the quantity represented by $\Gamma$ with the incoming states, and, on the other hand, an upper bound for such exchange for any normalized incoming state. 

We have recently shown that for any incoming state $\phiinket$ the exchange of $\Gamma$ can be written as (\cite[Eq. 3]{FerCor2016c})
\begin{equation}
	\DG=\phiinbra\Gamma-S^{\dagger}\Gamma S\phiinket,
\end{equation}
and split into the average value of two hermitian operators related to non-unitary interaction and asymmetry, respectively. This is an improvement over Noether's theorem because it allows to consider and quantify the exchange of $\Gamma$ in the presence of absorption or gain. If there is no absorption or gain, we can write (\cite[Eq. 3]{FerCor2016c})
\begin{equation}
\label{eq:dg}
	\DG=\phiinbra S^{\dagger}[S,\Gamma]\phiinket.
\end{equation}
We note that the above two equations are obtained in \cite{FerCor2016c} in the context of conservation laws for the electromagnetic field. It is nevertheless clear that their derivation does not depend on the nature of the incoming and outgoing states, and applies to the general scattering setting.

The norm of the operator $S^{\dagger}[S,\Gamma]$ in Eq. (\ref{eq:dg}) is then a quantity of interest. It provides an excitation independent indication of the ability of the system to exchange $\Gamma$ with the incoming states. It turns out that the Frobenius norm of $S^{\dagger}[S,\Gamma]$ is equal to the expression in Eq. (\ref{eq:limit2}):

\begin{equation}
	\label{eq:aiaiai}
	\begin{split}
		\normF{S^{\dagger}[S,\Gamma]}&=\sqrt{\Tr{\left(S^{\dagger}[S,\Gamma]\right)^\dagger\left(S^{\dagger}[S,\Gamma]\right)}}\\
																					 &=\sqrt{\Tr{[S,\Gamma]^\dagger SS^\dagger[S,\Gamma]}}\\
																	  &=\sqrt{\Tr{[S,\Gamma]^\dagger [S,\Gamma]}}\\&
		=\sqrt{B_{\Gamma}\left(4\Tr{S^\dagger S}\right)}=\sqrt{\sum\limits_{\gamma,\bargamma} \left(\gamma-\bargamma\right)^2\X},\\
	\end{split}
\end{equation}
where the third equality follows because $SS^\dagger$ is the identity when $S$ is unitary, and the fourth and fifth follow from Eqs. (\ref{eq:Msmall}) and (\ref{eq:Msmalllab}). 

Additionally, the performance of the most efficient electromagnetic field for transferring $\Gamma$ to the system can be bounded by Eq. (\ref{eq:limit2}). 
\begin{equation}
	\begin{split}
		&\biggl|\frac{\phiinbra S^{\dagger}[S,\Gamma]\phiinket}{\phiinbra\Phi_{\text{in}}\rangle}\biggr|\le \sigma_1\left(S^{\dagger}[S,\Gamma]\right)\le \normF{S^{\dagger}[S,\Gamma]}\\
		&=\sqrt{B_{\Gamma}\left(4\Tr{S^\dagger S}\right)}=\sqrt{\sum\limits_{\gamma,\bargamma} \left(\gamma-\bargamma\right)^2\X},
	\end{split}
\end{equation}
where $\sigma_1(A)$ is the largest singular value of $A$, the first inequality follows readily by using the singular value decomposition of $S^{\dagger}[S,\Gamma]$, the second one is a known inequality between different operator norms \cite[p. 17-6]{Hogben2006}, and the equalities follow again from Eqs. (\ref{eq:Msmall}) and (\ref{eq:Msmalllab}).

Table \ref{tab:C} shows the values of $C_{S \Gamma}$ for $\Gamma$ representing the angular momentum along the $\zhat$ axis $J_z$, and the performance achieved by the optimal monochromatic incoming state for inducing torque for the three systems in Fig. \ref{fig:breakings}.

\begin{table}[ht]
	\vspace{0.15cm}
\begin{tabular}{c c c c}
	& \hspace{0.1cm}\includegraphics[angle=90,width=0.6cm]{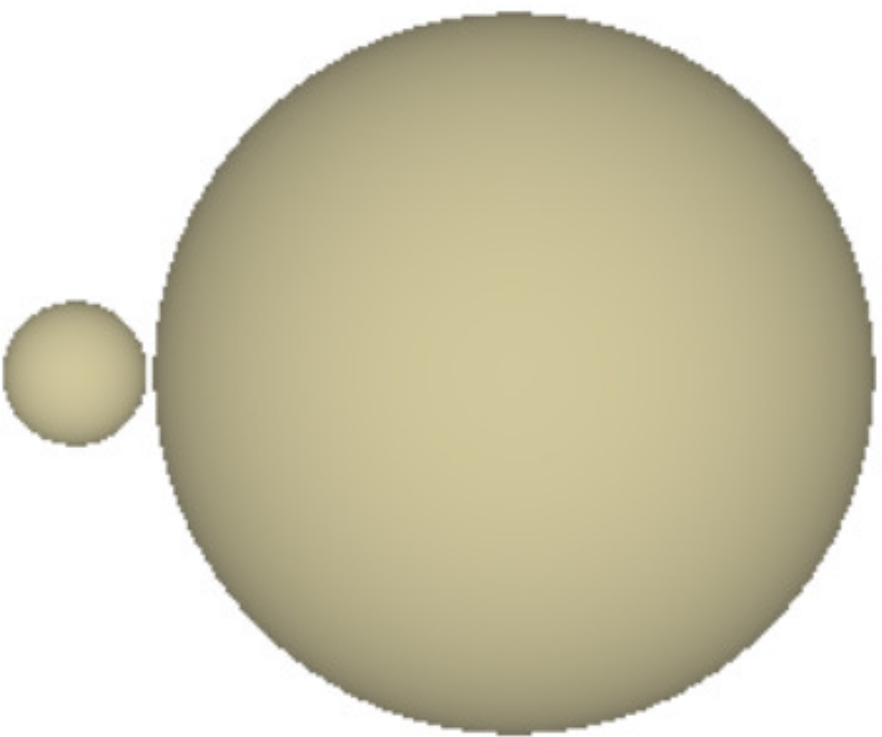} \hspace{0.1cm} &\hspace{0.1cm}\includegraphics[angle=90,width=0.8cm]{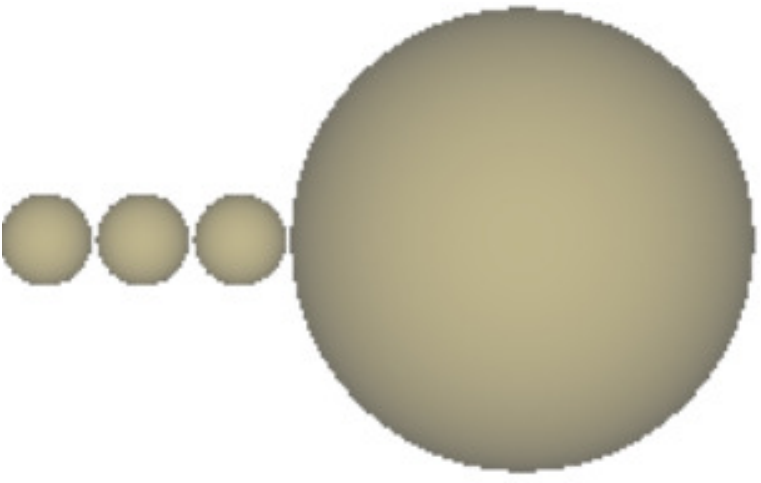} \hspace{0.1cm}& \hspace{0.1cm}\includegraphics[angle=90,width=0.6cm]{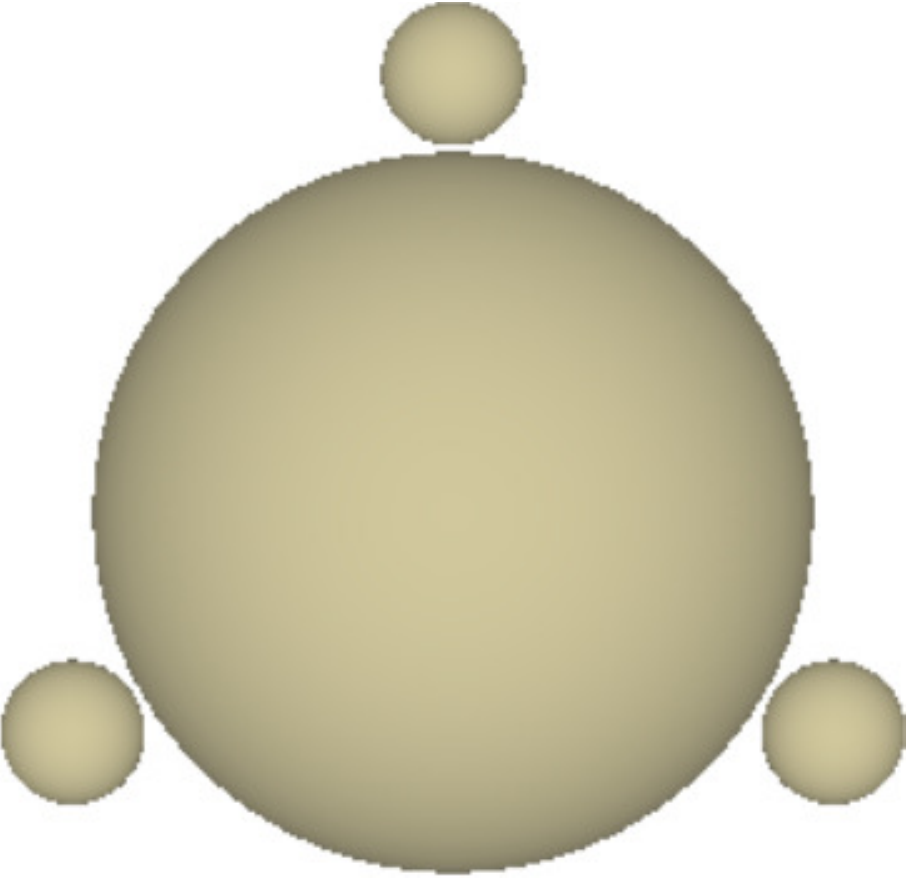} \hspace{0.1cm}\\
	$C_{S J_z}$ & 1.97 & 5.30 & 3.32\\
	Optimal & 0.80 & 2.51 & 0.99\\
\end{tabular}
\caption{\label{tab:C} The quantity $C_{S\Gamma}$ in Eq. (\ref{eq:limit}) bounds the optimal performance of any incoming state regarding transferring $\Gamma$ to the system. The table shows the values of the bound and the optimal transfer of angular momentum $J_z$ for the three systems in Fig. \ref{fig:breakings}.}
\end{table}

\section{$B_\Gamma=0\implies M[S,\Ttheta S {\Ttheta}^{-1}]=0\ \forall \ \theta$ \label{app:b0m0}}
We first rewrite the denominator of the symmetry breaking measurement for general $\theta$ in Eq. (\ref{eq:M}) by using the commutator between $S$ and $\Ttheta$: 
\begin{equation}
[S,\Ttheta]=S\Ttheta-\Ttheta S. 
\end{equation}

We note that 
\begin{equation}
	\begin{split}
		S-\Ttheta  S \Tmtheta&=S-\left\{S\Ttheta-[S,\Ttheta]\right\}\Tmtheta\\
							 &=[S,\Ttheta]\Tmtheta,
	\end{split}
\end{equation}
with which 
\begin{equation}
\label{eq:last}
	\begin{split}
		&\Tr{\left[S-\Ttheta  S \Tmtheta \right]^\dagger\left[S-\Ttheta  S \Tmtheta \right]}\\
		&=\Tr{\Ttheta[S,\Ttheta]^\dagger[S,\Ttheta]\Tmtheta}\\
		&=\Tr{[S,\Ttheta]^\dagger[S,\Ttheta]},
	\end{split}
\end{equation}
where the last equality follows from the cyclic property of the trace $\Tr{ABC}=\Tr{BCA}$.

The last line of Eq. (\ref{eq:last}) reflects the expected result that the symmetry breaking is zero if $S$ and $\Ttheta$ commute. We will finish proving the statement on this section's title by showing that $[S,\Gamma]=0\implies[S,\Ttheta]=0$, with which we conclude that if the slope of symmetry breaking $B_{\Gamma}$ vanishes, it means that all other higher order terms vanish as well.

In order to show that $[S,\Gamma]=0\implies [S,\Ttheta]=0$ we use that $\Ttheta=\exp\left(-i\theta\Gamma\right)=\sum_{l=0}^\infty \frac{(-i\theta\Gamma)^l}{l!}$ to write

\begin{equation}
		[S,\Ttheta]=\sum_{l=0}^{\infty}\frac{(-i\theta)^l}{l!}[S,\Gamma^l].
\end{equation}
We can write $[S,\Gamma^l]$ as a function of $[S,\Gamma]$ and $[S,\Gamma^{l-1}]$
\begin{equation}
	\label{eq:nyaca}
	\begin{split}
		[S,\Gamma^l]&=S\Gamma^l-\Gamma^lS=S\Gamma\Gamma^{l-1}-\Gamma\Gamma^{l-1} S\\
					&=\left\{[S,\Gamma]+\Gamma S\right\}\Gamma^{l-1}-\Gamma\Gamma^{l-1} S\\
					&=[S,\Gamma]\Gamma^{l-1}+\Gamma [S,\Gamma^{l-1}],
	\end{split}
\end{equation}
and finish by noting that $[S,\Gamma^0]=[S,I]=0$, and that $[S,\Gamma]=0$ by assumption, which then means that, for $l=2$, $[S,\Gamma^2]=0$ because of Eq. (\ref{eq:nyaca}), and so on for all $l$.

 \end{document}}{}
\begin{thebibliography}{45}\makeatletter
\providecommand \@ifxundefined [1]{ \@ifx{#1\undefined}
}\providecommand \@ifnum [1]{ \ifnum #1\expandafter \@firstoftwo
 \else \expandafter \@secondoftwo
 \fi
}\providecommand \@ifx [1]{ \ifx #1\expandafter \@firstoftwo
 \else \expandafter \@secondoftwo
 \fi
}\providecommand \natexlab [1]{#1}\providecommand \enquote  [1]{``#1''}\providecommand \bibnamefont  [1]{#1}\providecommand \bibfnamefont [1]{#1}\providecommand \citenamefont [1]{#1}\providecommand \href@noop [0]{\@secondoftwo}\providecommand \href [0]{\begingroup \@sanitize@url \@href}\providecommand \@href[1]{\@@startlink{#1}\@@href}\providecommand \@@href[1]{\endgroup#1\@@endlink}\providecommand \@sanitize@url [0]{\catcode `\\12\catcode `\$12\catcode
  `\&12\catcode `\#12\catcode `\^12\catcode `\_12\catcode `\%12\relax}\providecommand \@@startlink[1]{}\providecommand \@@endlink[0]{}\providecommand \url  [0]{\begingroup\@sanitize@url \@url }\providecommand \@url [1]{\endgroup\@href {#1}{\urlprefix }}\providecommand \urlprefix  [0]{URL }\providecommand \Eprint [0]{\href }\providecommand \doibase [0]{http://dx.doi.org/}\providecommand \selectlanguage [0]{\@gobble}\providecommand \bibinfo  [0]{\@secondoftwo}\providecommand \bibfield  [0]{\@secondoftwo}\providecommand \translation [1]{[#1]}\providecommand \BibitemOpen [0]{}\providecommand \bibitemStop [0]{}\providecommand \bibitemNoStop [0]{.\EOS\space}\providecommand \EOS [0]{\spacefactor3000\relax}\providecommand \BibitemShut  [1]{\csname bibitem#1\endcsname}\let\auto@bib@innerbib\@empty
\bibitem [{\citenamefont {Wigner}(1939)}]{Wigner1939}  \BibitemOpen
  \bibfield  {author} {\bibinfo {author} {\bibfnamefont {E.}~\bibnamefont
  {Wigner}},\ }\href {\doibase 10.2307/1968551} {\bibfield  {journal} {\bibinfo
   {journal} {Annals of Mathematics}\ }\textbf {\bibinfo {volume} {40}},\
  \bibinfo {pages} {149} (\bibinfo {year} {1939})},\ \bibinfo {note}
  {{ArticleType:} research-article / Full publication date: Jan., 1939 /
  Copyright © 1939 Annals of Mathematics}\BibitemShut {NoStop}\bibitem [{\citenamefont {Nambu}\ and\ \citenamefont
  {Jona-Lasinio}(1961)}]{Nambu1961}  \BibitemOpen
  \bibfield  {author} {\bibinfo {author} {\bibfnamefont {Y.}~\bibnamefont
  {Nambu}}\ and\ \bibinfo {author} {\bibfnamefont {G.}~\bibnamefont
  {Jona-Lasinio}},\ }\href {\doibase 10.1103/PhysRev.122.345} {\bibfield
  {journal} {\bibinfo  {journal} {Phys. Rev.}\ }\textbf {\bibinfo {volume}
  {122}},\ \bibinfo {pages} {345} (\bibinfo {year} {1961})}\BibitemShut
  {NoStop}\bibitem [{\citenamefont {Higgs}(1964)}]{Higgs1964}  \BibitemOpen
  \bibfield  {author} {\bibinfo {author} {\bibfnamefont {P.~W.}\ \bibnamefont
  {Higgs}},\ }\href {\doibase 10.1103/PhysRevLett.13.508} {\bibfield  {journal}
  {\bibinfo  {journal} {Phys. Rev. Lett.}\ }\textbf {\bibinfo {volume} {13}},\
  \bibinfo {pages} {508} (\bibinfo {year} {1964})}\BibitemShut {NoStop}\bibitem [{\citenamefont {Anderson}(1961)}]{Anderson1961}  \BibitemOpen
  \bibfield  {author} {\bibinfo {author} {\bibfnamefont {P.~W.}\ \bibnamefont
  {Anderson}},\ }\href {\doibase 10.1103/PhysRev.124.41} {\bibfield  {journal}
  {\bibinfo  {journal} {Phys. Rev.}\ }\textbf {\bibinfo {volume} {124}},\
  \bibinfo {pages} {41} (\bibinfo {year} {1961})}\BibitemShut {NoStop}\bibitem [{\citenamefont {Wilson}(1983)}]{Wilson1983}  \BibitemOpen
  \bibfield  {author} {\bibinfo {author} {\bibfnamefont {K.~G.}\ \bibnamefont
  {Wilson}},\ }\href {\doibase 10.1103/RevModPhys.55.583} {\bibfield  {journal}
  {\bibinfo  {journal} {Rev. Mod. Phys.}\ }\textbf {\bibinfo {volume} {55}},\
  \bibinfo {pages} {583} (\bibinfo {year} {1983})}\BibitemShut {NoStop}\bibitem [{\citenamefont {Penrose}(2017)}]{Penrose2017}  \BibitemOpen
  \bibfield  {author} {\bibinfo {author} {\bibfnamefont {R.}~\bibnamefont
  {Penrose}},\ }\href@noop {} {\bibfield  {journal} {\bibinfo  {journal} {arXiv
  preprint arXiv:1707.04169}\ } (\bibinfo {year} {2017})}\BibitemShut {NoStop}\bibitem [{\citenamefont {Wigner}(1959)}]{Wigner1959}  \BibitemOpen
  \bibfield  {author} {\bibinfo {author} {\bibfnamefont {E.~P.}\ \bibnamefont
  {Wigner}},\ }\href@noop {} {{\selectlanguage {english}\emph {\bibinfo {title}
  {Group Theory and Its Application to the Quantum Mechanics of Atomic
  Spectra}}}}\ (\bibinfo  {publisher} {Academic Press},\ \bibinfo {year}
  {1959})\BibitemShut {NoStop}\bibitem [{\citenamefont {Anderson}(1997)}]{Anderson1997}  \BibitemOpen
  \bibfield  {author} {\bibinfo {author} {\bibfnamefont {P.~W. P.~W.}\
  \bibnamefont {Anderson}},\ }\href@noop {} {{\selectlanguage {English}\emph
  {\bibinfo {title} {Basic notions of condensed matter physics}}}}\ (\bibinfo
  {publisher} {Reading, Mass. : Addison-Wesley},\ \bibinfo {year} {1997})\
  \bibinfo {note} {"The Advanced book program."}\BibitemShut {NoStop}\bibitem [{\citenamefont {Barron}(2004)}]{Barron2004}  \BibitemOpen
  \bibfield  {author} {\bibinfo {author} {\bibfnamefont {L.~D.}\ \bibnamefont
  {Barron}},\ }\href@noop {} {\emph {\bibinfo {title} {Molecular Light
  Scattering and Optical Activity}}},\ \bibinfo {edition} {2nd}\ ed.\ (\bibinfo
   {publisher} {Cambridge University Press},\ \bibinfo {year}
  {2004})\BibitemShut {NoStop}\bibitem [{\citenamefont {Noether}(1918)}]{Noether1918}  \BibitemOpen
  \bibfield  {author} {\bibinfo {author} {\bibfnamefont {E.}~\bibnamefont
  {Noether}},\ }\href@noop {} {\bibfield  {journal} {\bibinfo  {journal}
  {Nachr. v. d. Ges. d. Wiss. zu G\"ottingen}\ }\textbf {\bibinfo {volume}
  {1918}},\ \bibinfo {pages} {235} (\bibinfo {year} {1918})}\BibitemShut
  {NoStop}\bibitem [{\citenamefont {French}(1967)}]{French1967}  \BibitemOpen
  \bibfield  {author} {\bibinfo {author} {\bibfnamefont {J.}~\bibnamefont
  {French}},\ }\href {\doibase http://dx.doi.org/10.1016/0370-2693(67)90551-5}
  {\bibfield  {journal} {\bibinfo  {journal} {Physics Letters B}\ }\textbf
  {\bibinfo {volume} {26}},\ \bibinfo {pages} {75 } (\bibinfo {year}
  {1967})}\BibitemShut {NoStop}\bibitem [{\citenamefont {Brody}\ \emph {et~al.}(1981)\citenamefont {Brody},
  \citenamefont {Flores}, \citenamefont {French}, \citenamefont {Mello},
  \citenamefont {Pandey},\ and\ \citenamefont {Wong}}]{Brody1981}  \BibitemOpen
  \bibfield  {author} {\bibinfo {author} {\bibfnamefont {T.~A.}\ \bibnamefont
  {Brody}}, \bibinfo {author} {\bibfnamefont {J.}~\bibnamefont {Flores}},
  \bibinfo {author} {\bibfnamefont {J.~B.}\ \bibnamefont {French}}, \bibinfo
  {author} {\bibfnamefont {P.~A.}\ \bibnamefont {Mello}}, \bibinfo {author}
  {\bibfnamefont {A.}~\bibnamefont {Pandey}}, \ and\ \bibinfo {author}
  {\bibfnamefont {S.~S.~M.}\ \bibnamefont {Wong}},\ }\href {\doibase
  10.1103/RevModPhys.53.385} {\bibfield  {journal} {\bibinfo  {journal} {Rev.
  Mod. Phys.}\ }\textbf {\bibinfo {volume} {53}},\ \bibinfo {pages} {385}
  (\bibinfo {year} {1981})}\BibitemShut {NoStop}\bibitem [{\citenamefont {Wochner}\ \emph {et~al.}(2009)\citenamefont
  {Wochner}, \citenamefont {Gutt}, \citenamefont {Autenrieth}, \citenamefont
  {Demmer}, \citenamefont {Bugaev}, \citenamefont {Ortiz}, \citenamefont
  {Duri}, \citenamefont {Zontone}, \citenamefont {Grübel},\ and\ \citenamefont
  {Dosch}}]{Wochner2009}  \BibitemOpen
  \bibfield  {author} {\bibinfo {author} {\bibfnamefont {P.}~\bibnamefont
  {Wochner}}, \bibinfo {author} {\bibfnamefont {C.}~\bibnamefont {Gutt}},
  \bibinfo {author} {\bibfnamefont {T.}~\bibnamefont {Autenrieth}}, \bibinfo
  {author} {\bibfnamefont {T.}~\bibnamefont {Demmer}}, \bibinfo {author}
  {\bibfnamefont {V.}~\bibnamefont {Bugaev}}, \bibinfo {author} {\bibfnamefont
  {A.~D.}\ \bibnamefont {Ortiz}}, \bibinfo {author} {\bibfnamefont
  {A.}~\bibnamefont {Duri}}, \bibinfo {author} {\bibfnamefont {F.}~\bibnamefont
  {Zontone}}, \bibinfo {author} {\bibfnamefont {G.}~\bibnamefont {Grübel}}, \
  and\ \bibinfo {author} {\bibfnamefont {H.}~\bibnamefont {Dosch}},\ }\href
  {\doibase 10.1073/pnas.0905337106} {\bibfield  {journal} {\bibinfo  {journal}
  {Proceedings of the National Academy of Sciences}\ }\textbf {\bibinfo
  {volume} {106}},\ \bibinfo {pages} {11511} (\bibinfo {year}
  {2009})}\BibitemShut {NoStop}\bibitem [{\citenamefont {Reichert}\ \emph {et~al.}(2000)\citenamefont
  {Reichert}, \citenamefont {Klein}, \citenamefont {Dosch}, \citenamefont
  {Denk}, \citenamefont {Honkimaki}, \citenamefont {Lippmann},\ and\
  \citenamefont {Reiter}}]{Reichert2000}  \BibitemOpen
  \bibfield  {author} {\bibinfo {author} {\bibfnamefont {H.}~\bibnamefont
  {Reichert}}, \bibinfo {author} {\bibfnamefont {O.}~\bibnamefont {Klein}},
  \bibinfo {author} {\bibfnamefont {H.}~\bibnamefont {Dosch}}, \bibinfo
  {author} {\bibfnamefont {M.}~\bibnamefont {Denk}}, \bibinfo {author}
  {\bibfnamefont {V.}~\bibnamefont {Honkimaki}}, \bibinfo {author}
  {\bibfnamefont {T.}~\bibnamefont {Lippmann}}, \ and\ \bibinfo {author}
  {\bibfnamefont {G.}~\bibnamefont {Reiter}},\ }\href {\doibase
  10.1038/35048537} {\bibfield  {journal} {\bibinfo  {journal} {Nature}\
  }\textbf {\bibinfo {volume} {408}},\ \bibinfo {pages} {839} (\bibinfo {year}
  {2000})}\BibitemShut {NoStop}\bibitem [{\citenamefont {Marvian~Mashhad}(2012)}]{Marvian2012}  \BibitemOpen
  \bibfield  {author} {\bibinfo {author} {\bibfnamefont {I.}~\bibnamefont
  {Marvian~Mashhad}},\ }\emph {\bibinfo {title} {Symmetry, asymmetry and
  quantum information}},\ \href {http://hdl.handle.net/10012/7088} {Ph.D.
  thesis} (\bibinfo {year} {2012})\BibitemShut {NoStop}\bibitem [{\citenamefont {Marvian}\ and\ \citenamefont
  {Spekkens}(2014)}]{Marvian2014}  \BibitemOpen
  \bibfield  {author} {\bibinfo {author} {\bibfnamefont {I.}~\bibnamefont
  {Marvian}}\ and\ \bibinfo {author} {\bibfnamefont {R.~W.}\ \bibnamefont
  {Spekkens}},\ }\href {http://dx.doi.org/10.1038/ncomms4821} {\bibfield
  {journal} {\bibinfo  {journal} {Nat. Commun.}\ }\textbf {\bibinfo {volume}
  {5}},\ \bibinfo {pages} {3821} (\bibinfo {year} {2014})},\ \bibinfo {note}
  {article}\BibitemShut {NoStop}\bibitem [{\citenamefont {Fang}\ \emph {et~al.}(2016)\citenamefont {Fang},
  \citenamefont {Dong}, \citenamefont {Zhou},\ and\ \citenamefont
  {Sun}}]{Fang2016}  \BibitemOpen
  \bibfield  {author} {\bibinfo {author} {\bibfnamefont {Y.-N.}\ \bibnamefont
  {Fang}}, \bibinfo {author} {\bibfnamefont {G.-H.}\ \bibnamefont {Dong}},
  \bibinfo {author} {\bibfnamefont {D.-L.}\ \bibnamefont {Zhou}}, \ and\
  \bibinfo {author} {\bibfnamefont {C.-P.}\ \bibnamefont {Sun}},\ }\href
  {http://stacks.iop.org/0253-6102/65/i=4/a=423} {\bibfield  {journal}
  {\bibinfo  {journal} {Communications in Theoretical Physics}\ }\textbf
  {\bibinfo {volume} {65}},\ \bibinfo {pages} {423} (\bibinfo {year}
  {2016})}\BibitemShut {NoStop}\bibitem [{\citenamefont {Marvian}\ \emph {et~al.}(2016)\citenamefont
  {Marvian}, \citenamefont {Spekkens},\ and\ \citenamefont
  {Zanardi}}]{Marvian2016}  \BibitemOpen
  \bibfield  {author} {\bibinfo {author} {\bibfnamefont {I.}~\bibnamefont
  {Marvian}}, \bibinfo {author} {\bibfnamefont {R.~W.}\ \bibnamefont
  {Spekkens}}, \ and\ \bibinfo {author} {\bibfnamefont {P.}~\bibnamefont
  {Zanardi}},\ }\href {\doibase 10.1103/PhysRevA.93.052331} {\bibfield
  {journal} {\bibinfo  {journal} {Phys. Rev. A}\ }\textbf {\bibinfo {volume}
  {93}},\ \bibinfo {pages} {052331} (\bibinfo {year} {2016})}\BibitemShut
  {NoStop}\bibitem [{\citenamefont {Dong}\ \emph
  {et~al.}(2017{\natexlab{a}})\citenamefont {Dong}, \citenamefont {Fang},\ and\
  \citenamefont {Sun}}]{Dong2016}  \BibitemOpen
  \bibfield  {author} {\bibinfo {author} {\bibfnamefont {G.-H.}\ \bibnamefont
  {Dong}}, \bibinfo {author} {\bibfnamefont {Y.-N.}\ \bibnamefont {Fang}}, \
  and\ \bibinfo {author} {\bibfnamefont {C.-P.}\ \bibnamefont {Sun}},\ }\href
  {http://stacks.iop.org/0253-6102/68/i=4/a=405} {\bibfield  {journal}
  {\bibinfo  {journal} {Communications in Theoretical Physics}\ }\textbf
  {\bibinfo {volume} {68}},\ \bibinfo {pages} {405} (\bibinfo {year}
  {2017}{\natexlab{a}})}\BibitemShut {NoStop}\bibitem [{\citenamefont {Dong}\ \emph
  {et~al.}(2017{\natexlab{b}})\citenamefont {Dong}, \citenamefont {Zhang},
  \citenamefont {Sun},\ and\ \citenamefont {Gong}}]{Dong2017}  \BibitemOpen
  \bibfield  {author} {\bibinfo {author} {\bibfnamefont {G.~H.}\ \bibnamefont
  {Dong}}, \bibinfo {author} {\bibfnamefont {Z.~W.}\ \bibnamefont {Zhang}},
  \bibinfo {author} {\bibfnamefont {C.~P.}\ \bibnamefont {Sun}}, \ and\
  \bibinfo {author} {\bibfnamefont {Z.~R.}\ \bibnamefont {Gong}},\ }\href
  {\doibase 10.1038/s41598-017-13405-0} {\bibfield  {journal} {\bibinfo
  {journal} {Scientific Reports}\ }\textbf {\bibinfo {volume} {7}},\ \bibinfo
  {pages} {12947} (\bibinfo {year} {2017}{\natexlab{b}})}\BibitemShut {NoStop}\bibitem [{\citenamefont {Sorkin}(1994)}]{Sorkin1994}  \BibitemOpen
  \bibfield  {author} {\bibinfo {author} {\bibfnamefont {R.~D.}\ \bibnamefont
  {Sorkin}},\ }\href {\doibase 10.1142/S021773239400294X} {\bibfield  {journal}
  {\bibinfo  {journal} {Modern Physics Letters A}\ }\textbf {\bibinfo {volume}
  {09}},\ \bibinfo {pages} {3119} (\bibinfo {year} {1994})}\BibitemShut
  {NoStop}\bibitem [{\citenamefont {Hardy}(2001)}]{Hardy2001}  \BibitemOpen
  \bibfield  {author} {\bibinfo {author} {\bibfnamefont {L.}~\bibnamefont
  {Hardy}},\ }\href {https://arxiv.org/abs/quant-ph/0101012} {\bibfield
  {journal} {\bibinfo  {journal} {arXiv preprint quant-ph/0101012}\ } (\bibinfo
  {year} {2001})}\BibitemShut {NoStop}\bibitem [{\citenamefont {Reed}\ and\ \citenamefont {Simon}(1979)}]{Reed1979}  \BibitemOpen
  \bibfield  {author} {\bibinfo {author} {\bibfnamefont {M.}~\bibnamefont
  {Reed}}\ and\ \bibinfo {author} {\bibfnamefont {B.}~\bibnamefont {Simon}},\
  }\href@noop {} {\bibfield  {journal} {\bibinfo  {journal} {Scattering theory.
  Methods of modern mathematical physics, Vol. 3. New York, NY (USA): American
  Press.}\ } (\bibinfo {year} {1979})}\BibitemShut {NoStop}\bibitem [{\citenamefont {Colton}\ and\ \citenamefont
  {Kress}(2012)}]{Colton2012}  \BibitemOpen
  \bibfield  {author} {\bibinfo {author} {\bibfnamefont {D.}~\bibnamefont
  {Colton}}\ and\ \bibinfo {author} {\bibfnamefont {R.}~\bibnamefont {Kress}},\
  }\href@noop {} {\emph {\bibinfo {title} {Inverse acoustic and electromagnetic
  scattering theory}}},\ Vol.~\bibinfo {volume} {93}\ (\bibinfo  {publisher}
  {Springer Science \& Business Media},\ \bibinfo {address} {New York},\
  \bibinfo {year} {2012})\BibitemShut {NoStop}\bibitem [{\citenamefont {(auth.)}(1982)}]{Newton2013}  \BibitemOpen
  \bibfield  {author} {\bibinfo {author} {\bibfnamefont {R.~G.~N.}\
  \bibnamefont {(auth.)}},\ }\href@noop {} {\emph {\bibinfo {title} {Scattering
  Theory of Waves and Particles}}},\ \bibinfo {edition} {2nd}\ ed.,\ Texts and
  Monographs in Physics\ (\bibinfo  {publisher} {Springer-Verlag Berlin
  Heidelberg},\ \bibinfo {year} {1982})\BibitemShut {NoStop}\bibitem [{\citenamefont {Altland}\ and\ \citenamefont
  {Simons}(2010)}]{Altland2010}  \BibitemOpen
  \bibfield  {author} {\bibinfo {author} {\bibfnamefont {A.}~\bibnamefont
  {Altland}}\ and\ \bibinfo {author} {\bibfnamefont {B.~D.}\ \bibnamefont
  {Simons}},\ }\href@noop {} {{\selectlanguage {english}\emph {\bibinfo {title}
  {Condensed Matter Field Theory}}}}\ (\bibinfo  {publisher} {Cambridge
  University Press},\ \bibinfo {year} {2010})\BibitemShut {NoStop}\bibitem [{\citenamefont {Coleman}\ and\ \citenamefont
  {Mandula}(1967)}]{Coleman1967}  \BibitemOpen
  \bibfield  {author} {\bibinfo {author} {\bibfnamefont {S.}~\bibnamefont
  {Coleman}}\ and\ \bibinfo {author} {\bibfnamefont {J.}~\bibnamefont
  {Mandula}},\ }\href {\doibase 10.1103/PhysRev.159.1251} {\bibfield  {journal}
  {\bibinfo  {journal} {Phys. Rev.}\ }\textbf {\bibinfo {volume} {159}},\
  \bibinfo {pages} {1251} (\bibinfo {year} {1967})}\BibitemShut {NoStop}\bibitem [{\citenamefont {Weinberg}(1995)}]{Weinberg1995}  \BibitemOpen
  \bibfield  {author} {\bibinfo {author} {\bibfnamefont {S.}~\bibnamefont
  {Weinberg}},\ }\href
  {http://www.amazon.com/exec/obidos/redirect?tag=citeulike07-20\&path=ASIN/0521550017}
  {\emph {\bibinfo {title} {{The Quantum Theory of Fields (Volume 1)}}}},\
  \bibinfo {edition} {1st}\ ed.\ (\bibinfo  {publisher} {Cambridge University
  Press},\ \bibinfo {year} {1995})\BibitemShut {NoStop}\bibitem [{Note1()}]{Note1}  \BibitemOpen
  \bibinfo {note} {Even the choice of the Euclidean norm in the three-vector
  case above is arbitrary.}\BibitemShut {Stop}\bibitem [{Note2()}]{Note2}  \BibitemOpen
  \bibinfo {note} {If we know $S$, the choice of operator norm is not critical
  from the operational point of view because we can compute any operator norm
  we desire.}\BibitemShut {Stop}\bibitem [{Note3()}]{Note3}  \BibitemOpen
  \bibinfo {note} {This is a notable simplification with respect to the task of
  obtaining $S$ by experimental means. In general, both amplitude and phase
  measurements for complete sets of incoming states and outgoing measurements
  are needed to determine the complex elements of $S$. Phase measurements are
  known to be particularly challenging in many situations \protect \cite
  {Teague1983,Phillips2000}. $S$ may be analytically and/or numerically
  obtained when a model of the response of the system is available. Then, the
  experimentally obtained symmetry breaking measures allow to test the
  model.}\BibitemShut {Stop}\bibitem [{Note4()}]{Note4}  \BibitemOpen
  \bibinfo {note} {Then: ${T_{\Gamma }\left (\theta \right )}^{-1}=\left
  ({T_{\Gamma }\left (\theta \right )}\right )^{\protect \dagger }=\left
  [\protect \qopname \relax o{exp}\left (-i\theta \Gamma \right )\right
  ]^{\protect \dagger }=\protect \qopname \relax o{exp}\left (i\theta \Gamma
  \right )=T_{\Gamma }\left (-\theta \right )$}\BibitemShut {NoStop}\bibitem [{Note5()}]{Note5}  \BibitemOpen
  \bibinfo {note} {It is worth noting that the number of $\Gamma $ eigenvalues
  can be infinite. For example, when considering rotations, $\gamma $ and
  $\protect \mathaccentV {bar}016{\gamma }$ can take any integer value. In
  theory this implies that an infinite number of $X_{\protect \mathaccentV
  {bar}016{\gamma }\gamma }$ has to be obtained. In practice, the response of a
  system of finite size will eventually fall rapidly as the modulo of the
  angular momentum eigenvalue increases. This provides a limit to the number of
  needed $X_{\protect \mathaccentV {bar}016{\gamma }\gamma }$ for the rotation
  case. Similar arguments can be found in other cases.}\BibitemShut {Stop}\bibitem [{\citenamefont {Fernandez-Corbaton}\ and\ \citenamefont
  {Rockstuhl}(2017)}]{FerCor2016c}  \BibitemOpen
  \bibfield  {author} {\bibinfo {author} {\bibfnamefont {I.}~\bibnamefont
  {Fernandez-Corbaton}}\ and\ \bibinfo {author} {\bibfnamefont
  {C.}~\bibnamefont {Rockstuhl}},\ }\href {\doibase 10.1103/PhysRevA.95.053829}
  {\bibfield  {journal} {\bibinfo  {journal} {Phys. Rev. A}\ }\textbf {\bibinfo
  {volume} {95}},\ \bibinfo {pages} {053829} (\bibinfo {year}
  {2017})}\BibitemShut {NoStop}\bibitem [{\citenamefont {Kosteleck\'y}(2004)}]{Kosteleck2004}  \BibitemOpen
  \bibfield  {author} {\bibinfo {author} {\bibfnamefont {V.~A.}\ \bibnamefont
  {Kosteleck\'y}},\ }\href {\doibase 10.1103/PhysRevD.69.105009} {\bibfield
  {journal} {\bibinfo  {journal} {Phys. Rev. D}\ }\textbf {\bibinfo {volume}
  {69}},\ \bibinfo {pages} {105009} (\bibinfo {year} {2004})}\BibitemShut
  {NoStop}\bibitem [{\citenamefont {Mattingly}(2005)}]{Mattingly2005}  \BibitemOpen
  \bibfield  {author} {\bibinfo {author} {\bibfnamefont {D.}~\bibnamefont
  {Mattingly}},\ }\href {\doibase 10.12942/lrr-2005-5} {\bibfield  {journal}
  {\bibinfo  {journal} {Living Reviews in Relativity}\ }\textbf {\bibinfo
  {volume} {8}},\ \bibinfo {pages} {5} (\bibinfo {year} {2005})}\BibitemShut
  {NoStop}\bibitem [{\citenamefont {Liberati}(2013)}]{Liberati2013}  \BibitemOpen
  \bibfield  {author} {\bibinfo {author} {\bibfnamefont {S.}~\bibnamefont
  {Liberati}},\ }\href {http://stacks.iop.org/0264-9381/30/i=13/a=133001}
  {\bibfield  {journal} {\bibinfo  {journal} {Classical and Quantum Gravity}\
  }\textbf {\bibinfo {volume} {30}},\ \bibinfo {pages} {133001} (\bibinfo
  {year} {2013})}\BibitemShut {NoStop}\bibitem [{\citenamefont {Teague}(1983)}]{Teague1983}  \BibitemOpen
  \bibfield  {author} {\bibinfo {author} {\bibfnamefont {M.~R.}\ \bibnamefont
  {Teague}},\ }\href {\doibase 10.1364/JOSA.73.001434} {\bibfield  {journal}
  {\bibinfo  {journal} {J. Opt. Soc. Am.}\ }\textbf {\bibinfo {volume} {73}},\
  \bibinfo {pages} {1434} (\bibinfo {year} {1983})}\BibitemShut {NoStop}\bibitem [{\citenamefont {Phillips}\ \emph {et~al.}(2000)\citenamefont
  {Phillips}, \citenamefont {Knight}, \citenamefont {Pottage}, \citenamefont
  {Kakarantzas},\ and\ \citenamefont {Russell}}]{Phillips2000}  \BibitemOpen
  \bibfield  {author} {\bibinfo {author} {\bibfnamefont {P.~L.}\ \bibnamefont
  {Phillips}}, \bibinfo {author} {\bibfnamefont {J.~C.}\ \bibnamefont
  {Knight}}, \bibinfo {author} {\bibfnamefont {J.~M.}\ \bibnamefont {Pottage}},
  \bibinfo {author} {\bibfnamefont {G.}~\bibnamefont {Kakarantzas}}, \ and\
  \bibinfo {author} {\bibfnamefont {P.~S.~J.}\ \bibnamefont {Russell}},\ }\href
  {\doibase 10.1063/1.125812} {\bibfield  {journal} {\bibinfo  {journal}
  {Applied Physics Letters}\ }\textbf {\bibinfo {volume} {76}},\ \bibinfo
  {pages} {541} (\bibinfo {year} {2000})}\BibitemShut {NoStop}\bibitem [{\citenamefont {Tischler}\ \emph {et~al.}(2014)\citenamefont
  {Tischler}, \citenamefont {Fernandez-Corbaton}, \citenamefont
  {Zambrana-Puyalto}, \citenamefont {Minovich}, \citenamefont {Vidal},
  \citenamefont {Juan},\ and\ \citenamefont {Molina-Terriza}}]{Tischler2014}  \BibitemOpen
  \bibfield  {author} {\bibinfo {author} {\bibfnamefont {N.}~\bibnamefont
  {Tischler}}, \bibinfo {author} {\bibfnamefont {I.}~\bibnamefont
  {Fernandez-Corbaton}}, \bibinfo {author} {\bibfnamefont {X.}~\bibnamefont
  {Zambrana-Puyalto}}, \bibinfo {author} {\bibfnamefont {A.}~\bibnamefont
  {Minovich}}, \bibinfo {author} {\bibfnamefont {X.}~\bibnamefont {Vidal}},
  \bibinfo {author} {\bibfnamefont {M.~L.}\ \bibnamefont {Juan}}, \ and\
  \bibinfo {author} {\bibfnamefont {G.}~\bibnamefont {Molina-Terriza}},\ }\href
  {\doibase 10.1038/lsa.2014.64} {\bibfield  {journal} {\bibinfo  {journal}
  {Light Sci. Appl.}\ }\textbf {\bibinfo {volume} {3}},\ \bibinfo {pages}
  {e183} (\bibinfo {year} {2014})}\BibitemShut {NoStop}\bibitem [{\citenamefont {Lavery}\ \emph {et~al.}(2013)\citenamefont {Lavery},
  \citenamefont {Robertson}, \citenamefont {Sponselli}, \citenamefont
  {Courtial}, \citenamefont {Steinhoff}, \citenamefont {Tyler}, \citenamefont
  {Willner},\ and\ \citenamefont {Padgett}}]{Lavery2013}  \BibitemOpen
  \bibfield  {author} {\bibinfo {author} {\bibfnamefont {M.~P.~J.}\
  \bibnamefont {Lavery}}, \bibinfo {author} {\bibfnamefont {D.~J.}\
  \bibnamefont {Robertson}}, \bibinfo {author} {\bibfnamefont {A.}~\bibnamefont
  {Sponselli}}, \bibinfo {author} {\bibfnamefont {J.}~\bibnamefont {Courtial}},
  \bibinfo {author} {\bibfnamefont {N.~K.}\ \bibnamefont {Steinhoff}}, \bibinfo
  {author} {\bibfnamefont {G.~A.}\ \bibnamefont {Tyler}}, \bibinfo {author}
  {\bibfnamefont {A.}~\bibnamefont {Willner}}, \ and\ \bibinfo {author}
  {\bibfnamefont {M.~J.}\ \bibnamefont {Padgett}},\ }\href
  {http://stacks.iop.org/1367-2630/15/i=1/a=013024} {\bibfield  {journal}
  {\bibinfo  {journal} {New Journal of Physics}\ }\textbf {\bibinfo {volume}
  {15}},\ \bibinfo {pages} {013024} (\bibinfo {year} {2013})}\BibitemShut
  {NoStop}\bibitem [{\citenamefont {Ren}\ \emph {et~al.}(2016)\citenamefont {Ren},
  \citenamefont {Li}, \citenamefont {Zhang},\ and\ \citenamefont
  {Gu}}]{Ren2016}  \BibitemOpen
  \bibfield  {author} {\bibinfo {author} {\bibfnamefont {H.}~\bibnamefont
  {Ren}}, \bibinfo {author} {\bibfnamefont {X.}~\bibnamefont {Li}}, \bibinfo
  {author} {\bibfnamefont {Q.}~\bibnamefont {Zhang}}, \ and\ \bibinfo {author}
  {\bibfnamefont {M.}~\bibnamefont {Gu}},\ }\href {\doibase
  10.1126/science.aaf1112} {\bibfield  {journal} {\bibinfo  {journal}
  {Science}\ }\textbf {\bibinfo {volume} {352}},\ \bibinfo {pages} {805}
  (\bibinfo {year} {2016})}\BibitemShut {NoStop}\bibitem [{\citenamefont {Xu}(1995)}]{Xu1995}  \BibitemOpen
  \bibfield  {author} {\bibinfo {author} {\bibfnamefont {Y.-l.}\ \bibnamefont
  {Xu}},\ }\href {\doibase 10.1364/AO.34.004573} {\bibfield  {journal}
  {\bibinfo  {journal} {Appl. Opt.}\ }\textbf {\bibinfo {volume} {34}},\
  \bibinfo {pages} {4573} (\bibinfo {year} {1995})}\BibitemShut {NoStop}\bibitem [{\citenamefont {Tung}(1985)}]{Tung1985}  \BibitemOpen
  \bibfield  {author} {\bibinfo {author} {\bibfnamefont {W.-K.}\ \bibnamefont
  {Tung}},\ }\href@noop {} {\emph {\bibinfo {title} {Group Theory in
  Physics}}}\ (\bibinfo  {publisher} {World Scientific},\ \bibinfo {year}
  {1985})\BibitemShut {NoStop}\bibitem [{\citenamefont {Hogben}(2006)}]{Hogben2006}  \BibitemOpen
  \bibfield  {author} {\bibinfo {author} {\bibfnamefont {L.}~\bibnamefont
  {Hogben}},\ }\href@noop {} {\emph {\bibinfo {title} {Handbook of linear
  algebra}}}\ (\bibinfo  {publisher} {CRC Press},\ \bibinfo {year}
  {2006})\BibitemShut {NoStop}\end{thebibliography}
